\documentclass[11pt,a4paper]{article}

\usepackage{jheppub}
\usepackage[utf8]{inputenc} 
\usepackage[normalem]{ulem}
\usepackage{graphicx}
\usepackage{dcolumn}
\usepackage{enumerate}

\usepackage{mathrsfs} 
\usepackage[dvipsnames,table]{xcolor}
\usepackage{cancel}
\usepackage{bbold}
\usepackage{braket}
\usepackage{physics}
\usepackage{multirow}
\usepackage[capitalize]{cleveref}
\usepackage{xspace}
\usepackage{comment}
\usepackage{fontawesome} 
\definecolor{nicegreen}{rgb}{0., 0.75, 0.46}

\definecolor{wildstrawberry}{rgb}{1.0, 0.26, 0.64}
\newcommand{\pjf}[1]{{\color{wildstrawberry}\bf PJF: #1}}

\definecolor{MH}{rgb}{0.0,0.6,9}
\definecolor{palatinate}{rgb}{0.494, 0.192, 0.482}
\definecolor{blue-violet}{rgb}{0.33, 0.17, 0.89}

\renewcommand{\phi}{\varphi}
\def\ltap{\ \raise.3ex\hbox{$<$\kern-.75em\lower1ex\hbox{$\sim$}}\ }
\def\gtap{\ \raise.3ex\hbox{$>$\kern-.75em\lower1ex\hbox{$\sim$}}\ }
\def\lsim{\ \raise.3ex\hbox{$<$\kern-.75em\lower1ex\hbox{$\sim$}}\ }
\def\gsim{\ \raise.3ex\hbox{$>$\kern-.75em\lower1ex\hbox{$\sim$}}\ }

\newcommand{\be}{\begin{equation}}
\newcommand{\ee}{\end{equation}}
\newcommand{\beq}{\begin{equation}}
\newcommand{\eeq}{\end{equation}}
\newcommand{\bea}{\begin{eqnarray}}
\newcommand{\eea}{\end{eqnarray}}
\newcommand{\bear}{\begin{eqnarray}}
\newcommand{\eear}{\end{eqnarray}}

\DeclareMathOperator{\br}{BR}

\def\MeV{\,{\rm MeV}}

\newcommand{\e}{{\rm e}}

\preprint{FERMILAB-PUB-25-0818-T}
\preprint{MI-HET-869}
\preprint{CALT-TH/2025-034}
\preprint{CERN-TH-2025-229}

\title{$L_\mu-L_\tau$ gauge bosons in beam dumps and supernovae}

\author[a]{Nikita Blinov,}
\author[b]{Patrick J.\ Fox,}
\author[c]{Kevin J.\ Kelly,}
\author[d,e]{Ryan Plestid}
\author[c,f]{and Tao Zhou}

\affiliation[a]{Department of Physics and Astronomy, York University,\\
Toronto, Ontario, M3J 1P3, Canada}

\affiliation[b]{Particle Theory Department, Fermi National Accelerator Laboratory,\\
Batavia, IL 60510, USA}

\affiliation[c]{Department of Physics and Astronomy, Mitchell Institute for Fundamental Physics and Astronomy,\\
Texas A\&M University, College Station, TX 77843, USA}

\affiliation[d]{Walter Burke Institute for Theoretical Physics,\\
California Institute of Technology, Pasadena, CA 91125, USA}

\affiliation[e]{Theoretical Physics Department, CERN,\\
1 Esplanade des Particules, CH-1211 Geneva 23, Switzerland}

\affiliation[f]{Department of Physics, University of Cincinnati,\\
Cincinnati, OH 45221, USA}

\emailAdd{nblinov@yorku.ca}
\emailAdd{pjfox@fnal.gov}
\emailAdd{kjkelly@tamu.edu}
\emailAdd{ryan.plestid@cern.ch}
\emailAdd{taozhou@tamu.edu}

\abstract{
    We study the phenomenology of a sub-GeV $L_\mu-L_\tau$ gauge boson. We find discrepancies with existing literature in sensitivity projections for the upcoming SHiP experiment and in the treatment of supernovae cooling constraints. We present a quantitative analysis of different production modes in beam dumps and compare our results to previous work. In the context of supernovae, we re-evaluate the standard supernova cooling bounds from SN1987A and analyze additional supernova-based probes: diffusive cooling, constraints from the existence of low-energy supernovae, and the absence of a high-energy neutrino signal from SN1987A.}

\begin{document}
\maketitle

\section{Introduction \label{sec:Introduction} }
New particles with masses well below the weak scale may exist if they have sufficiently feeble couplings to ordinary matter \cite{Antel:2023hkf}. They are naturally motivated by the existence of dark matter, and are therefore often assumed to be part of a larger ``dark sector'' \cite{Alexander:2016aln}. When these particles' masses lie in the MeV-GeV range, they are most efficiently searched for in high intensity fixed target experiments~\cite{Beacham:2019nyx,Ilten:2022lfq}. 

One of the most experimentally challenging models of light new physics involves a mediator which couples exclusively to second-  and third-generation leptons~\cite{Bauer:2018onh}. A theoretically well motivated example of this type occurs when the anomaly-free flavor symmetry of the Standard Model $L_\mu-L_\tau$~\cite{Foot:1990mn,He:1990pn, He:1991qd,Heeck:2011wj} is gauged. When the mass of the gauge boson $V$ lies below the di-muon threshold, the branching ratio for $V\rightarrow \nu \bar{\nu}$ is nearly unity, with only a small $\sim 10^{-5}$ probability of producing a $e^+e^-$ final state. This makes searches for visible signals of $L_\mu-L_\tau$ gauge bosons very challenging.\!\footnote{We refer the interested reader to Ref.~\cite{Bauer:2018onh} for a thorough summary of searches for $L_\mu - L_\tau$ gauge bosons in laboratory experiments.} In contrast, related models like $L_e-L_\mu$ or $L_e-L_\tau$ (which have tree-level couplings to electrons) are subject to a similar multitude of constraints as the thoroughly-studied dark photon~\cite{Bauer:2018onh,Ilten:2018crw}.

Muonphillic and electrophobic force carriers are further motivated by anomalies in the muon sector~\cite{Baek:2001kca, Ma:2001md}, and serve as a challenging target for more general dark sector searches. Astrophysical environments, such as the muon- and neutrino-rich core of supernovae and neutron stars~\cite{Escudero:2019gzq,Croon:2020lrf,Cerdeno:2023kqo, Akita:2023iwq, Lai:2024mse}, also offer complementary sensitivity to laboratory searches. 
The distinctive phenomenology of $L_\mu-L_\tau$ thus motivates studying it independently of other leptonic gauge symmetries like $L_e-L_\mu$ and $L_e-L_\tau$. Here, we examine its signatures in both fixed-target experiments and supernova environments.
\begin{figure}[t]
    \centering
    \includegraphics[width=0.9\linewidth]{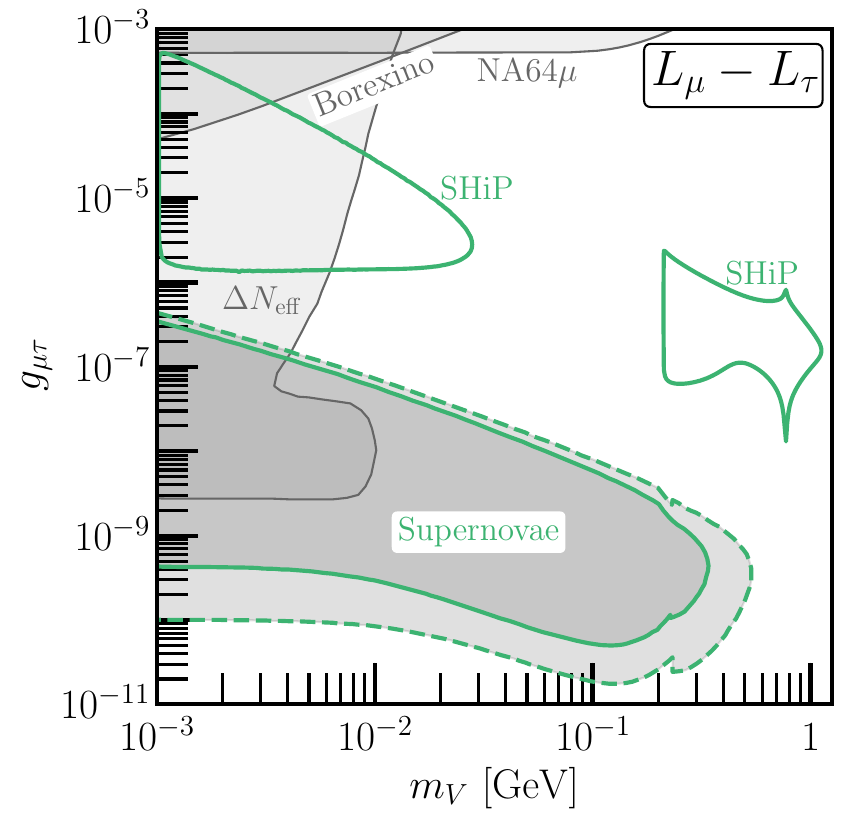}
    \caption{Constraints and experimental projections in the mass-coupling plane for the $L_\mu - L_\tau$ gauge boson. We derive new reach estimates for the SHiP beam-dump experiment, assuming that $6\times 10^{20}$ protons on target will be collected (green solid lines). We also show an updated combination of supernovae constraints (shaded gray region with dashed and solid green lines corresponding to large/small progenitor masses, respectively). Previous constraints from early-universe measurements of $\Delta N_{\rm eff}$~\cite{Escudero:2019gzq}, neutrino-electron scattering in Borexino~\cite{Altmannshofer:2019zhy,Kelly:2024tvh}, from NA64$\mu$~\cite{NA64:2024klw} are shown for comparison.\label{fig:summary}}
\end{figure}

The search for $L_{\mu}-L_{\tau}$ at the upcoming SHiP fixed-target experiment~\cite{Aberle:2839677,Albanese:2878604} exhibits several qualitative differences from a generic dark photon or electrophilic gauge bosons. For the $L_{\mu}-L_{\tau}$ model, production through neutral meson decays and electromagnetic secondaries is strongly suppressed by the loop-induced nature of the kinetic mixing with the photon. This means that processes which make use of tree-level couplings to the dark vector, e.g., charged meson decays and muon bremsstrahlung~\cite{Chen:2017awl,Amaral:2021rzw,Cesarotti:2023udo,Cesarotti:2023sje} may be important. We have re-evaluated the sensitivity of SHiP to the $L_{\mu}-L_{\tau}$ model, including production from tree-level and loop-induced interactions. Our findings are summarized in~\cref{fig:summary}. Despite the loop suppression, we find that meson decays make the largest contribution to the $V$-flux below the dimuon threshold. In this parameter space, the signal rates are suppressed by the small branching fraction to $e^+e^-$. Above the dimuon threshold, the visible branching fraction is significant; there meson decays and proton bremsstrahlung contribute to the flux. We find that production through tree-level couplings is typically subdominant: charged meson decays suffer from phase-space suppression, while flux from muon bremsstrahlung is small for several reasons discussed in \cref{sec:beam-dumps}. We have also included $V$-production from electromagnetic cascades, following~\cite{Zhou:2024aeu}; these processes can produce large fluxes, but with energies below the nominal thresholds of the detector. The contributions of different processes to the detector flux are shown in \cref{fig:flux10MeV} for $m_V = 10\;\MeV$. 

Turning to non-laboratory-based constraints, we have updated past analyses of supernovae cooling~\cite{Escudero:2019gzq,Croon:2020lrf, Lai:2024mse} in several ways. First we re-evaluated certain thermally-averaged production rates, ensuring they satisfy detailed balance conditions. Second, we incorporated several missing effects which have been studied recently~\cite{Caputo:2022rca}: stimulated emission as $V$'s propagate out of the supernova, and the redshift of the $V$-luminosity due to the large gravitational potential. We find that this more careful treatment leads to cooling constraints (as computed via the Raffelt criterion~\cite{Raffelt:1996wa}) that differ by factors of at most three in the coupling, and qualitatively agree with past literature. 

In addition to cooling bounds, we also analyze several other constraints arising from supernovae observations. These include bounds from diffusive energy loss~\cite{Cerdeno:2023kqo}, the non-observation of high energy ($\gtrsim 100~{\rm MeV}$) neutrinos from SN1987A~\cite{Akita:2022etk,Fiorillo:2022cdq,Akita:2023iwq}, from supernova progenitor heating by decays to charged final states, e.g., $V\to \mu^+\mu^-$~\cite{Caputo:2022mah} and from $\gamma$ rays emitted in radiative decays $V\to e^+e^- \gamma,\;\mu^+\mu^-\gamma$~\cite{Caputo:2021rux}. To the best of our knowledge this is the first comprehensive treatment of all of these effects for the $L_\mu-L_\tau$ model. The combination of all of the supernova bounds is shown in~\cref{fig:summary}.

The rest of the paper is organized as follows. In \cref{sec:model} we define the $L_\mu-L_\tau$ model. Updated SHiP results and a detailed analysis of the relevant production mechanisms are presented in \cref{sec:beam-dumps}. Constraints from supernovae are discussed in \cref{sec:supernovae}, and conclusions are drawn in \cref{sec:conclusions}. \cref{sec:beamdump_appendix} offers more detail regarding our SHiP projections, while \cref{sec:sn_appendix} provides additional details of the supernova calculations. 

\section{Model definition \& preliminaries \label{sec:model} }
The Standard Model (SM) contains a set of accidental symmetries related to lepton flavor ($L_e$, $L_\mu$, and $L_\tau$) and baryon number ($B$). 
Combinations of the form $L_i-L_j$ are anomaly-free, and thus can be gauged, 
without extending the SM's matter content.
It is therefore possible that some of the accidental symmetries of the Standard Model are in fact exact and descend from a gauge symmetry.
In what follows we will be specifically interested in $L_\mu-L_\tau$. 

The gauge boson can become massive, either via the Higgs or Stuckelberg mechanism.
At low energies, such a theory is described by the effective Lagrangian, 
\begin{equation}
    \label{eq:eff-L}
    \begin{split}
    \mathcal{L}\supset -\frac{1}{4} \mathcal{V}_{\mu\nu} \mathcal{V}^{\mu\nu} + \frac12 m_V^2 V^{\mu} V_\mu  + g_{\mu\tau} J^\mu V_\mu~. 
    \end{split}
\end{equation}
where  $\mathcal{V}_{\mu\nu} = \partial_\mu V_\nu - \partial_\nu V_\mu$ is the field strength tensor of $U(1)_{L_\mu-L_\tau}$ and 
\begin{equation}
    J^\mu=\sum_f Q_f' \bar{f} \gamma^\mu f~,
\end{equation}
and the sum runs over fermions in the Standard Model with each fermion's charge under $L_{\mu}-L_{\tau}$ given by $Q_f'$. 
Loops of $\mu$ and $\tau$ pairs lead to a finite, calculable kinetic mixing with the photon~\cite{Kamada:2015era}:
\begin{align}
    \!\!\!\varepsilon(q^2) & =\! -\frac{g_{\mu\tau} e}{2\pi^2} \!\!\int_0^1\!\! \dd x~x(1-x) \log\qty[\frac{m_\tau^2-x(1-x)q^2}{m_\mu^2-x(1-x)q^2}] \label{eq:kinetic_mixing} \\
    & \sim 10^{-6} \left(\frac{g_{\mu\tau}}{10^{-4}}\right).\nonumber
\end{align}
This means that an off-shell photon in any amplitude can be replaced with a $V$ by inserting a factor of $\varepsilon(q^2)$.
The mixing then mediates interactions of $V$ with quarks and hadrons. 
In principle, heavy states that have been integrated out of the theory may supply a ``bare'' kinetic mixing in the effective Lagrangian \cref{eq:eff-L}, but we include only the irreducible contribution from $\mu$ and $\tau$ lepton loops in what follows. 

An important difference between $L_\mu-L_\tau$ and, e.g., a dark photon, is that the current $J^\mu$ contains $\nu_\mu$ and $\nu_\tau$, but {\it does not} contain the electron since $Q_e'=0$.
It is naturally neutrinophilic and electrophobic.
This qualitatively changes the experimental reach of beam dump experiments and constraints from supernovae compared to scenarios with tree-level. 

\section{Phenomenology in beam dumps \label{sec:beam-dumps} }
In this section we focus on the beam-dump phenomenology of $U(1)_{L_\mu-L_\tau}$. This model has the following properties which make it challenging to constrain by searching for displaced decays of $V$ to visible final states:
\begin{itemize}
    \item $V$-production rates are controlled either by the small kinetic mixing with the Standard Model photon, \cref{eq:kinetic_mixing}, or by reactions involving muons.\!\footnote{If the beam energy is high enough, taus can be produced from $D$-meson decays, but the rate is small enough to be ignored.} $V$-production from secondary muons is also suppressed.
    \item For $m_V \leq 2 m_\mu$, the branching fraction for visible decays $\br(V\to e^+e^-) \sim \varepsilon^2 / g_{\mu\tau}^2 \sim 10^{-5}$ is tiny. 
    \item With larger couplings (and greater production), the decay length becomes too short to reach the downstream decay pipe. 
\end{itemize}
Alternative search strategies involve looking for missing energy in electron~\cite{NA64:2022rme} or muon beam experiments~\cite{NA64:2024klw} as $V$ decays invisibly $V\to \bar{\nu}\nu$, or in modifications of neutrino scattering through virtual $V$-exchange~\cite{Altmannshofer:2019zhy,Kelly:2024tvh}. The former approach enjoys a favorable scaling with the small coupling to the dark sector, since one only needs to produce the $V$, which has an $\mathcal{O}(1)$ invisible branching fraction. Unfortunately, missing energy experiments need to precisely measure the beam momentum before interaction, which limits their total exposure. In contrast, neutrino scattering experiments benefit from enormous exposures to solar neutrinos, but the rates of $V$-mediated neutrino scattering scale with the fourth power of the dark sector coupling and pay the kinetic mixing penalty. In practice, this means searches for visibly-decaying long-lived particles are the only means of exploring smaller $L_\mu-L_\tau$ couplings.

Below we consider a number of production mechanisms for the $L_\mu-L_\tau$ gauge boson arising from both tree-level and loop-induced couplings to SM particles. We analyze the resulting detector fluxes and use them to derive new sensitivity projections for SHiP. The latter differ substantially from previous estimates. 
Our main findings are: {\it i)} that the sensitivity of SHiP has been overestimated in the past, with the largest discrepancy in mass reach below the dimuon threshold, {\it ii)} that SHiP's mass reach is largely dominated by neutral meson decays (i.e., $\pi^0 \to \gamma V$), and {\it iii)} that the reach increases appreciably in both mass and coupling with increased exposures. We provide a detailed comparison of our results with previous literature in~\cref{sec:beamdump_appendix}. 
\begin{figure*}[!htbp]
    \centering
    \includegraphics[width=0.65\linewidth]{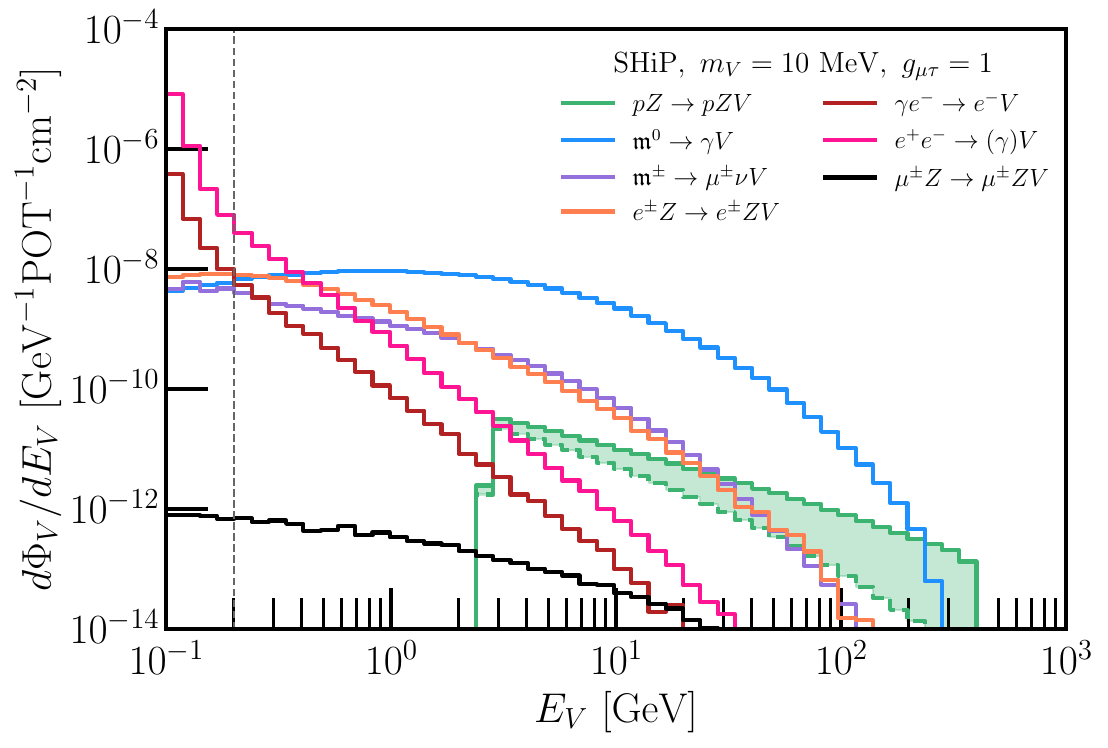}
    \caption{Flux of $L_{\mu}-L_\tau$ dark vectors pointed at the SHiP spectrometer prior to their decay. We have chosen the mass $m_V=10~{\rm MeV}$ for illustration, since many of the flux components contribute. The dashed vertical line at $E_V = 200$~MeV represents the energy cut that we assume for LLP reconstruction in the SHiP spectrometer. For proton bremsstrahlung (green), we shade the region between the expected flux obtained for different choices of a proton form factor parameter $\Lambda_p$, with $\Lambda_p = 1$~GeV (dashed) and $\Lambda_p\to \infty$ (solid); see text and~\cref{sec:beamdump_appendix} for further discussion.}
     \label{fig:flux10MeV}
\end{figure*}

\textbf{Production Mechanisms at SHiP:}
SHiP is a proton beam-dump facility sited at CERN.
It will collide the $400$~GeV SPS beam on a thick target made of ${\sim}60$ cm of molybdenum (corresponding to $\sim 4$ nuclear interaction lengths or 60 radiation lengths), followed by ${\sim}80$ cm tungsten~\cite{LopezSola:2019sfp}.  The target is followed by a 5~m long, magnetized iron hadron absorber. 
The HSDS (Hidden Sector Decay Spectrometer) features a $50$-m-long decay chamber that provides SHiP with strong discovery potential in the small-coupling/long-lived-particle regime. 
The majority of the beam interactions and secondary particle production will occur at the beginning of the Mo, so we study interactions in a uniform molybdenum target for simplicity.
In the $L_\mu - L_\tau$ model, the leading production channels include
\begin{itemize}
    \item Neutral meson ($\pi^0$, $\eta$, $\eta^\prime$)
    decay, $\mathfrak{m} \to \gamma V$, and $\omega \to \pi^0 V$,
    \item Charged meson ($\pi^\pm$, $K^\pm$)
decay, $\mathfrak{m}^\pm \to \ell^\pm \nu V$,
    \item Proton bremsstrahlung $p Z \to p Z V$, 
    \item Electron/positron bremsstrahlung, $e^\pm Z \to e^\pm Z V$.
\end{itemize}
We have found that the following channels are generally subleading:
\begin{itemize}
    \item Positron annihilation, $e^+ e^- \to (\gamma)V$,
    \item Compton scattering, $\gamma e^- \to e^- V$,
    \item Muon bremsstrahlung, $\mu^\pm Z \to \mu^\pm Z$.
\end{itemize}
We show the contributions of these processes to the total $V$-flux for a single parameter point in \cref{fig:flux10MeV}. Below we detail our predictions for these production channels.

We simulate the production of neutral and charged mesons using \texttt{GEANT-4}~\cite{GEANT4:2002zbu,Allison:2006ve,Allison:2016lfl} to model the interactions of a 400 GeV beam with a cylindrical Mo target with a radius 125 mm and length 60 cm~\cite{LopezSola:2019sfp}. 
Neutral mesons are used both to simulate direct $V$ production via $\mathfrak{m} \to \gamma V$, and as the initial `seed' for electromagnetic cascades in the Mo from $\mathfrak{m} \to \gamma\gamma$.
The high-energy nature of the SPS leads to very high-energy cascades, enabling very large electron/positron/photon multiplicities above energy thresholds necessary for $V$ production.
Our simulation of electromagnetic production of dark gauge bosons with \texttt{PETITE}~\cite{Blinov:2024pza} (via electron/positron bremsstrahlung, positron annihilation, and Compton scattering) is described in detail in Ref.~\cite{Zhou:2024aeu}; the only modification here is that such production is penalized by the loop-induced kinetic mixing from \cref{eq:kinetic_mixing}, with typical values $\varepsilon^2 \approx 10^{-5}g_{\mu\tau}^2$.

Previous work had utilized \texttt{PYTHIA-8}~\cite{SHiP:2020vbd,Zhou:2024aeu} or tabulated~\cite{Aguilar-Benitez:1991hzq} meson production differential cross sections at $400$~GeV incident proton energy~\cite{Bauer:2018onh}.
\texttt{PYTHIA-8}-based simulations yield conservative estimates of LLP production, as they focus on the primary proton-proton interactions and the neutral mesons (and subsequent electromagnetic cascade) from them.
We find that replacing a \texttt{PYTHIA-8}-based simulation with a \texttt{GEANT-4}-based one increases overall LLP production by approximately a factor of two, especially for lower-energy LLPs (as the bulk of secondary particle production accounted for in \texttt{GEANT-4} occurs at lower energies).
This has a minor, but visible, impact on our resulting sensitivity.
In comparing with results using tabulated meson-production differential cross sections, we find that our sensitivity estimates are considerably weaker; we provide further comparison and discussion of these differences in~\cref{sec:beamdump_appendix}.

\begin{figure}[t]
    \centering
    \includegraphics[width=0.9\linewidth]{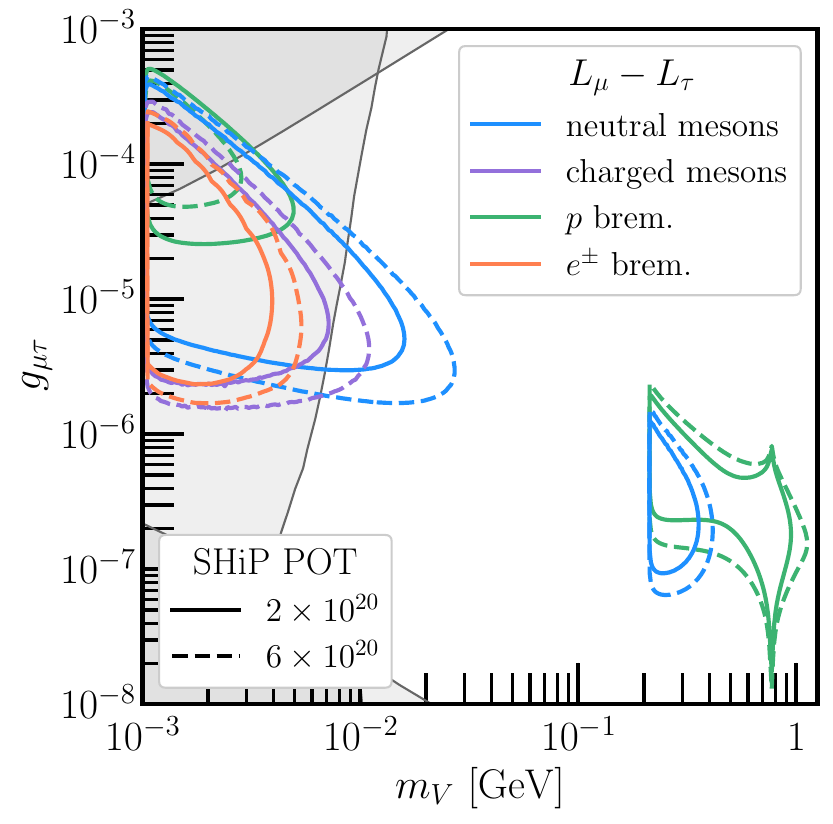}
    \caption{Sensitivity of SHiP with $2\times 10^{20}$ (solid) or $6\times 10^{20}$ (dashed) protons on target. In the parameter space that SHiP is able to test, and that is unconstrained by cosmology, production is dominated by neutral mesons produced by the primary proton beam and/or proton bremsstrahlung (above the dimuon threshold). 
    The low-mass sensitivity is governed by $V\to e^+ e^-$ decays, where above $m_V = 2 m_\mu$, decays into $\mu^+ \mu^-$ dictate the sensitivity.
    For proton bremsstrahlung, we take $\Lambda_p \to \infty$ for demonstration. Comparisons with past literature results and the impact of $\Lambda_p$ are discussed in~\cref{sec:beamdump_appendix}.
    \label{fig:SHiP_Sensitivity}}
\end{figure}

Charged mesons, also coming from \texttt{GEANT-4}, are utilized for direct decays and for secondary production from daughter muons. 
The decay channels include $\pi^{\pm}, K^{\pm} \rightarrow e^{\pm} \nu_e V$ and $\pi^{\pm}, K^{\pm} \rightarrow \mu^{\pm} \nu_\mu V$; however, the former suffer from both chirality~\cite{Krnjaic:2019rsv} and loop suppression (as the $V$ is radiated from $e^\pm$). 
Therefore, we only consider the muon channels. We also require that the decays occur within the first meson interaction length to avoid accounting for further attenuation in the target or deflection by magnetic fields in the hadron absorber.
Importantly, while the production rates of mesons themselves are fixed by QCD independently of the BSM physics, the relative role of neutral and charged mesons in vector production depends on the specific coupling of the model.
In the $L_{\mu}-L_{\tau}$ case charged-meson decays can make a significant contribution to the signal rate at lower vector masses due to their tree-level couplings to muons and neutrinos.
This is in contrast to the dark photon scenario~\cite{Zhou:2024aeu} where neutral mesons are typically the overwhelming source of production across all masses. Despite the large multiplicity of neutral mesons, their contribution to $L_\mu-L_\tau$ production is suppressed by the loop-induced kinetic mixing with the photon. These effects can be seen in \cref{fig:SHiP_Sensitivity} where the reach from individual channels is presented.

We also simulate the propagation of the muons, emerging from kaon/pion decay, through the Mo target.
This allows us to study muon bremsstrahlung of the gauge boson $\mu^\pm Z \to \mu^\pm Z V$, which has been implemented in \texttt{PETITE}~\cite{Blinov:2024pza}.
We find that, despite the tree-level coupling of the vector to the charged muons, this production mechanism is subdominant for the mass range of interest for SHiP.

Finally, for proton bremsstrahlung, we adopt the splitting function for $p \to pV$ emission discussed in Eq.~(12) of Ref.~\cite{Foroughi-Abari:2024xlj} using the quasi-real approximation (QRA), intended to improve upon previous estimates using the modified Williams-Weizs\"{a}cker (WW) approximation~\cite{Blumlein:2013cua,deNiverville:2016rqh,Berryman:2019dme}.
This approach incorporates two form factors: one for the electromagnetic properties of the proton (leading to enhancement when $m_V$ is close to the SM $\rho$/$\omega$ masses), and one to treat the off-shell nature of the proton involved in the splitting/emission of $V$.
We include a simplified form of the electromagnetic form factor to account for the $\omega$ and $\rho$ mesons of the SM that is slightly less rich than the form factor used in~\cite{Foroughi-Abari:2024xlj} but captures the main physics.
For the off-shell impact, we use the same structure as~\cite{Foroughi-Abari:2024xlj}, implementing a dipole form factor governed by a scale $\Lambda_p$. 
We find that the inclusion of this form factor and the specific choice of $\Lambda_p$ have a substantial impact on the experimental sensitivity using proton bremsstrahlung; we explore this in much more detail in~\cref{sec:beamdump_appendix}. For the remainder of the main text, we take $\Lambda_p \to \infty$ so that this form factor does not have any effect.

\textbf{Flux at SHiP:} The produced $L_\mu - L_\tau$ gauge boson flux for a dark vector mass of $10$~MeV for these channels is shown in~\cref{fig:flux10MeV} (we include only the flux pointed at the downstream SHiP spectrometer). The geometry of SHiP and the probability of decay products hitting the active detector volume is handled following the treatment in Ref.~\cite{Zhou:2024aeu}. This includes the updates to SHiP's geometry from Ref.~\cite{Albanese:2878604}. 
Neutral meson decays (blue), such as $\pi^0 \rightarrow \gamma V$, provide the largest overall contribution, dominating much of the flux over a broad energy range.
Charged meson decays (purple) provide a nontrivial contribution as well.
Resonant production from $e^+e^- \rightarrow V$ (pink) exhibits a sharp peak due to the resonance at $E_V = m_V^2/2m_e$. 
However, for this dark vector mass, the resonant energy sits below the energy threshold $E_V > 200$~MeV we impose for reconstruction in the SHiP spectrometer.
Instead, production from $e^\pm$ bremsstrahlung (orange) dominates among the electromagnetic channels.

For proton bremsstrahlung (green), we show the impact of varying $\Lambda_p$ for the proton off-shell form factor between $1$~GeV (dashed) and $\Lambda_p \to \infty$ (solid). The difference at $m_V = 10$~MeV is already large, and the impact is much greater for larger $V$ masses ${\sim}1$~GeV. We explore the large impact on experimental sensitivity from the specific choice of $\Lambda_p$ in~\cref{sec:beamdump_appendix} and hope that this spurs future phenomenological and experimental work to further hone the QRA formalism.

Production via muon bremsstrahlung (black) is small for a number of reasons: (1) the majority of charged mesons are absorbed in the target or hadron absorber before they can decay, which reduces the parent population that could yield muons -- as a result, far fewer muons traverse the target than electrons; (2) muons that do emerge traverse the target with minimal interaction (the muon path is typically under 1 radiation length), so they radiate far less than electrons; (3) the two-body kinematics of charged-meson decays fix the muon to be monoenergetic and isotropic in the meson rest frame, this maps into a box-shaped muon energy distribution after boosting to the lab frame. While the boost collimates most muons forward, SHiP's angular acceptance allows only the high-energy, small angle tail of this distribution, limiting the number of muons capable of emitting dark vectors within the detector.
Taken together, these effects render muon bremsstrahlung a negligible source of $L_\mu - L_\tau$ gauge bosons for all masses of interest at SHiP, despite the muon's tree-level coupling to $V$\@.

\textbf{Sensitivity at SHiP:} Following the approach detailed in Ref.~\cite{Zhou:2024aeu}, we calculate expected event rates in SHiP under different expected exposure scenarios for each of these different production channels.
\cref{fig:SHiP_Sensitivity} displays the resulting sensitivity as a function of the vector mass and the $L_\mu - L_\tau$ gauge coupling.
For the exposures considered, the only channels that lead to reach beyond existing bounds are neutral meson decay, charged meson decay, proton bremsstrahlung, and electron/positron bremsstrahlung.
Another notable feature in~\cref{fig:SHiP_Sensitivity} is that the reach contours consist of two disconnected islands, above and below the dimuon threshold at $m_V=2m_\mu\approx0.21$ GeV. For lighter masses, the vector boson cannot decay to $\mu^+\mu^-$; its visible signatures are restricted to loop-induced decays to $e^+e^-$. In this regime, the experimental reach is relatively weak compared to electrophilic gauge bosons due to the small branching fraction. Above the dimuon threshold, the branching ratio to $\mu^+\mu^-$ quickly dominates among visible channels. This leads to a substantial improvement in sensitivity. In addition, at $m_V\approx800~\textrm{MeV}$, the reach is enhanced by $\rho$ and $\omega$ resonance in proton bremsstrahlung~\cite{Blumlein:2013cua,Foroughi-Abari:2024xlj}. 

SHiP projections for $L_\mu-L_\tau$ have appeared in the literature before~\cite{Bauer:2018onh}. These calculations relied on rescaling of existing dark photon results. We compare our first-principles calculation with past literature in \cref{sec:beamdump_appendix}, generally finding that the estimated reach of the first-principles calculation is substantially weaker than the recast of the dark photon results.

\section{Supernovae constraints \label{sec:supernovae} }
Core-collapse supernovae (SN) contain a quantum-degenerate proto-neutron star (PNS), surrounded by a hot and dense stellar progenitor. 
High temperatures and chemical potentials ensure that neutrinos and muons are abundant in these environments. This means that $L_\mu - L_\tau$ boson production is possible in the SN via its tree-level interactions. 

Observations of supernovae can be used to constrain new physics in several ways, which we discuss below. The ${\sim}10$ second duration of the observed supernova 1987A (SN1987A) neutrino burst~\cite{Kamiokande-II:1987idp,Bionta:1987qt,Alekseev:1988gp} is consistent with SM expectations.
This time-scale is determined by the slow diffusive transport of neutrinos out of the PNS towards the neutrinosphere at $R_\nu\sim 25~{\rm km}$, where the neutrinos begin to free-stream. 
Light dark-sector states with sizable couplings to SM particles can rapidly cool supernovae, shortening the neutrino burst~\cite{Raffelt:1996wa}.
This is the origin of the so-called Raffelt or cooling bound: light new physics cannot carry away more than the neutrino luminosity. Applying this requirement at one second after core bounce yields a luminosity bound of 
\begin{equation}
    L_{R} = 4.4\times 10^{52}\; \mathrm{erg/s}
    \label{eq:raffelt_lumi}
\end{equation}
for the SN models used here~\cite{Caputo:2022rca}. 
Any energy-loss mechanism exceeding this, $L_{V}(g_{\mu\tau},m_V)\geq L_{R}$, is excluded by the observed SN1987A neutrino burst duration. 

Although the cooling argument is straightforward, the avenues for energy loss are not.
It has been recently shown that in the analogous case of cooling from axion (or axion-like particle) emission, there are many different regimes of energy transport out of the PNS~\cite{Caputo:2022mah,Fiorillo:2025yzf}.
In principle, there are two relevant regimes for a light $L_\mu-L_\tau$ gauge boson: ballistic and diffusive energy loss.
We discuss each of these in \cref{sec:ballistic} and \cref{sec:diffusive}, respectively, emphasizing discrepancies with existing literature.
While cooling bounds on $L_\mu-L_\tau$ have been evaluated before~\cite{Escudero:2019gzq,Croon:2020lrf,Lai:2024mse}, we implement several improvements in the calculation of the $V$-emission rates and the resulting luminosity $L_V$.
The improvements include: corrections to production rates; full angular averaging over emitted boson direction; finite-$m_V$ corrections in the optical depth of the SN medium; stimulated emission and gravitational effects.
These are described in detail in \cref{sec:sn_appendix}.

Additional constraints can be derived by comparing predictions of the high-energy neutrino flux to the SN1987A data.
A light $L_\mu-L_\tau$ boson can be produced in a high-$T$ region inside the PNS and decay outside, generating a flux of neutrinos with greater energy than the standard diffusive process can produce.
The absence of these neutrinos in observations provides further constraints on dark-sector states.
This bound was first calculated in the context of $L_\mu-L_\tau$ in ~\cite{Akita:2023iwq}; we re-evaluate it in \cref{sec:high_E_nu} using our improved $V$-luminosity described above.

Finally, for $m_V\geq 2m_\mu$, there is a sizable rate of $V\rightarrow \mu^+\mu^-$ decays which take place outside the neutrinosphere but inside the progenitor star.\!\footnote{Below the dimuon threshold there is a small rate of $V\rightarrow e^+e^-$ but the $\sim 10^{-5}$ branching ratio is small enough to make this channel irrelevant.}
These decays contribute to the explosion energy.
Existence of low-energy supernova (${\sim}10$ times dimmer than a typical core-collapse SN) can place a stringent constraint on such visibly-decaying particles~\cite{Falk:1978kf,Sung:2019xie,Caputo:2021rux,Caputo:2022mah,Fiorillo:2025sln,Fiorillo:2025yzf}.
We apply this argument to $L_\mu-L_\tau$ in~\cref{sec:low_energy_sn}. 

Let us also briefly note that another class of SN constraints can be derived from photons produced by dark sector particles emitted in SN~\cite{Caputo:2021rux}. These photons can contribute to the gamma-ray flux from SN1987A, or to the diffuse background from all past supernovae. In the $L_\mu-L_\tau$ model, photons can be produced in radiative decays $V\to e^+e^-\gamma,\;\mu^+\mu^- \gamma$. We estimate $\gamma$ ray emission in \cref{sec:sn_gamma_ray} following \cite{Caputo:2021rux}, and find that it does not provide a constraint on the $L_\mu-L_\tau$ parameter space. X-ray photons can also be produced if a thermalized region of electromagnetic plasma forms outside of a SN~\cite{Diamond:2023scc,Candon:2025ypl} or NS-NS merger~\cite{Diamond:2021ekg,Diamond:2023cto}. We discuss the formation of such a fireball in \cref{sec:fireballs} and estimate that this effect is unlikely to constrain new parameter space beyond neutrino scattering bounds. 

In what follows, we treat neutrinos as in thermal equilibrium inside the PNS allowing for different chemical potentials for the neutrino flavors.
We do not consider flavor instabilities (that would distort the $\nu_\mu$, $\nu_e$, and/or $\nu_\tau$ spectra) in our analysis; however we expect these effects to be small.
The proto-neutron star is sufficiently dense and hot that the neutrino gas should be isotropic to a very high level of accuracy.
Moreover, recently identified ``slow instabilities'' primarily operate outside the neutrinosphere~\cite{Fiorillo:2025gkw}, while ``fast instabilities'' require angular crossings, and do not play an important role in the first ten seconds post-bounce~\cite{Johns:2025mlm}.
Even if neutrino instabilities were to dramatically alter the density of either $\nu_\mu \bar{\nu}_\mu$ or  $\nu_\tau \bar{\nu}_\tau$ pairs, the effect would introduce at most a factor of two change in the overall luminosity.

We now turn to discussing each of the supernova constraints described above in turn, before presenting our results in \cref{fig:combined_sn_bounds}.

\subsection{Ballistic energy loss \label{sec:ballistic} }
The simplest irreducible energy loss mechanism is the production of $V$ at $r < R_\nu$, followed by decay at $r > R_\nu$.
For $L_\mu-L_\tau$ in the relevant parameter space, the $V$ decays to neutrinos which free stream outwards. This is the standard loss mechanism used to constrain many light dark sector models; here we refer to it as ballistic loss to distinguish it from diffusive loss (discussed in \cref{sec:diffusive}).
The luminosity due to this process is 
\begin{equation}
    \label{eq:ballistic-loss}
    L_{V,\rm ballistic} = \bigg\langle \omega ~\Gamma_{\rm prod}(\vb{r},\omega) P_{\rm esc}(\vb{r},\omega) \bigg\rangle ~,
\end{equation}
with $\omega$ the energy carried by $V$ at spatial infinity.
The average is carried out over phase space $(\int \dd^3r ~\dd^3 k/(2\pi)^3)$ inside the neutrinosphere, where $\Gamma_{\rm prod}$ is the rate of production per unit volume of phase space (see \cref{eq:full_luminosity} for a precise definition of $L_{V,\rm ballistic}$).  This is weighted by the probability that the $V$ escapes the supernova, $P_{\rm esc}(\vb{r},\omega)$, which includes thermal, geometric, and gravitational effects. The dominant production channels for $L_\mu - L_\tau$ bosons are inverse neutrino decays $\bar{\nu}_{\mu,\tau} \nu_{\mu,\tau} \to V$, Compton-like scattering $\gamma \mu \to V \mu$ and inverse muon decay $\bar{\mu} \mu \to V$, with bremsstrahlung-type reactions ($\mu p \to \mu p +V$~\cite{Croon:2020lrf}, $p n \to p n + V$~\cite{Lai:2024mse} and $\nu n \to \nu n + V$) being highly sub-dominant. The time-reversed versions of these processes contribute to absorption and therefore determine $P_{\rm esc}$.

We have improved on existing bounds by: accounting for neutrino chemical potentials in the inverse decay process, correcting small errors in production rates; performing a full average over the directions of $V$-emission; including gravitational time-dilation and redshift effects; using self-consistent definitions of production and absorption rates. These issues are described in detail in \cref{sec:sn_appendix}.

We have evaluated \cref{eq:ballistic-loss} numerically including the above improvements for two different 1D SN profiles (\texttt{s18.8-SFHo-MUONS} and \texttt{s20.0-SFHo-MUONS}~\cite{Mirizzi:2015eza,Bollig:2017lki,Bollig:2020xdr}) generated by the \textsc{Prometheus-Vertex} codes.\!\footnote{See \href{https://wwwmpa.mpa-garching.mpg.de/ccsnarchive}{wwwmpa.mpa-garching.mpg.de/ccsnarchive} for details.}  These models include the production of muon in the PNS and they correspond to progenitor masses of $18.8 M_\odot$ and $20M_\odot$ respectively, which roughly spans the allowed mass range of the SN1987A progenitor. The resulting constraints are shown in \cref{fig:combined_sn_bounds} by solid black lines. 
The higher-progenitor-mass model results in a hotter core and stronger sensitivity to dark sector particles.
The differences between these two models are used as an estimate of the systematic uncertainty in the derived bounds as in Refs.~\cite{Bollig:2020xdr,Croon:2020lrf}.
The new effects incorporated result in $\mathcal{O}(1)$ shifts to the energy loss but do not substantially alter the shape or location of supernovae exclusion limits computed in the literature.

\subsection{Diffusive energy loss \label{sec:diffusive}}
At large couplings, the mean free path of $V$ becomes very short.
Since $P_{\rm esc}$ is exponentially sensitive to these quantities, ballistic energy loss becomes inefficient.
Nevertheless, in this ``strongly-interacting'' regime an enormous number of dark vectors are produced, and the energy loss can still be substantially affected because of modified energy transport within the supernova.
This has been recently discussed in the context of axions~\cite{Caputo:2022rca,Caputo:2022mah,Fiorillo:2025sln,Fiorillo:2025yzf}.

As pointed out in~\cite{Cerdeno:2023kqo}, in the presence of an unstable $V$ that decays predominantly to $\nu\bar{\nu}$-pairs, neutrinos can effectively diffuse faster than is possible in the Standard Model. This is achieved by ``converting'' neutrinos into $V$s which propagate a distance of order their mean free path, before decaying back into $\nu \bar{\nu}$.
The mean free path $\lambda = v_V/\Gamma_{\rm abs}^*$ is determined by $v_V$, the vector velocity and by $\Gamma_{\rm abs}^*$, the reduced absorption rate defined in \cref{sec:sn_appendix} ($\Gamma_{\rm abs}^*$ also determines $P_{\rm esc}$ in \cref{eq:ballistic-loss}); these quantities depend on the energy of vector and the position inside the PNS. In the parameter space where $V$-diffusion is important, $\lambda$ is mainly determined by the in-medium decay length $V\to \bar{\nu} \nu$ (other processes like inverse Compton absorption $\mu V \to \mu \gamma$ and $V\to\mu^+\mu^-$ are subdominant).
As a rough estimate of $\lambda$, one may take the decay length for a dark vector $V$ with energy $\omega \sim T$: 
\begin{equation}
    \lambda \sim 1.5~{\rm km} \qty(\frac{\omega}{50 ~{\rm MeV}}) \qty(\frac{m_V}{5 ~{\rm MeV}})^{-2} \qty(\frac{10^{-7}}{g_{\mu\tau}})^2~.
\end{equation}
We see that for $g_{\mu\tau}$ close to the ballistic energy-loss ceiling (c.f. \cref{fig:combined_sn_bounds}), the $V$ mean free path is much smaller than the PNS radius $\sim R_\nu \approx 25$ km.

The comparison above indicates that for $g_{\mu\tau} \gtrsim 10^{-7}$ the PNS becomes a tightly-coupled fluid of $V$'s and neutrinos on scales larger than $\lambda$, which modifies the diffusive transport of neutrinos out of the PNS. 
In the absence of a dark vector, the mean free path of neutrinos in supernovae is $\sim 1~{\rm km}$ (this is a temperature-, and therefore radius-dependent statement).
In the presence of $L_\mu-L_\tau$ gauge bosons, inverse decays $\bar{\nu} \nu \rightarrow V$ dominate over Standard Model charged-current interactions for $g_{\mu\tau}\gtrsim 10^{-10}$. 
As a result, the mean free path of neutrinos is effectively set by the $V$ mean free path $\lambda$. 

We must now turn this observation into an estimate for energy loss and apply the Raffelt criterion, \cref{eq:raffelt_lumi}.
The Boltzmann equation for the $V$-distribution in the plane-parallel approximation can be solved perturbatively in $\lambda/R$ around the 
local equilibrium distribution $f_{\rm eq}(\vb{p},T)$~\cite{Caputo:2022rca}, where $R \sim R_\nu$. 
The result is 
\begin{equation}
    f(\vb{p},r) = f_{\rm eq}(\vb{p},T) - \lambda \cos\theta \pdv{T}{r} \pdv{T} f_{\rm eq}(\vb{p},T) + \ldots~,
    \label{eq:phase_space_dist_diffusion}
\end{equation}
where $\vb{p}$ is the $V$-momentum, $T=T(r)$ is the temperature profile and $\cos\theta= \hat{\vb{p}}\cdot \hat{\vb{r}}$. 
The rate of energy flux through a spherical shell of radius $R$ is then given by the phase space integral of this distribution weighed by $\vb{p}\cdot\hat{\vb{r}}$:
\begin{equation}
    \frac{\dd E}{\dd t} = - (4\pi R^2)\times \frac{1}{6\pi^2 T^2} \pdv{T}{r}  \int \dd \omega  \lambda  \frac{\vb{p}^2 \omega^2 \e^{-\omega/T}}{(1-\e^{-\omega/T})^2}~,
\end{equation}
where all quantities are evaluated at $r=R$,  and $\omega$ is the energy of $V$, respectively. 
We compare this flux to the Raffelt criterion and show the resulting bounds in \cref{fig:combined_sn_bounds} as dashed blue lines. 

The diffusive energy loss depends on the radius $R$. This radius must be chosen to satisfy several conditions which constrain the applicability of the previous calculation. First, we need to have $R \leq R_\nu$, since our thermal rate calculations assume that neutrinos are in local thermal equilibrium. Second, in order for the typical $V$ produced near $R$ to contribute to cooling of the PNS, it must have a reasonable probability to escape beyond $R_\nu$. This happens when $\lambda(R) > R_\nu - R$. Finally, the diffusion approximation is valid only when $\lambda(R) < R$ (that is, the higher-order terms in \cref{eq:phase_space_dist_diffusion} can be dropped). Together these conditions require that 
\beq
R > \lambda(R) > R_\nu - R.
\label{eq:diffusion_validity}
\eeq
In other words, the emission occurs from a thin shell of thickness $R_\nu - R$ at radius $R$. At a given $m_V$ and fixed $R$, this inequality restricts the range of $g_{\mu\tau}$ over which the diffusive calculation is appropriate. We find that $R_\nu - R = 5$ km gives a non-negligible region of parameter space where \cref{eq:diffusion_validity} is satisfied and the diffusive Raffelt bound holds. In \cref{fig:combined_sn_bounds} the diffusion bounds terminate where \cref{eq:diffusion_validity} fails to be satisfied (specifically, $\lambda(R)$ exceeds $R$). Unfortunately, and in contrast to \cite{Cerdeno:2023kqo}, we find that the diffusion argument fails to constrain parameter space beyond the regular ballistic cooling bound. We note that evaluating the diffusion at $R = R_\nu$ trivializes one of the conditions in \cref{eq:diffusion_validity}, but the resulting bound fails $\lambda < R_\nu$ so we do not show it.

\subsection{High-energy neutrinos}
\label{sec:high_E_nu}
Next, let us consider an alternative method for constraining parameter space in neutrinophilic models, which is based on searching for signals of high-energy neutrinos from supernovae. This idea was first proposed in the context of Majorons \cite{Akita:2022etk,Fiorillo:2022cdq} and then applied to heavy neutral leptons \cite{Syvolap:2023trc}, dark photons with di-muon final states \cite{Syvolap:2024hdh}, and recently to an $L_\mu-L_\tau$ gauge boson~\cite{Akita:2023iwq}. SN1987A data showed no evidence of neutrinos above $\sim 75~{\rm MeV}$~\cite{Kamiokande-II:1987idp,Bionta:1987qt,Alekseev:1988gp}. If there is a new cooling channel, such as from a long-lived particle decaying to neutrinos, $V\rightarrow \nu \bar{\nu}$, then decays outside the neutrinosphere will produce an additional flux of energetic neutrinos that can be observed in terrestrial detectors. If event rates were simply proportional to the neutrino flux, then this method would yield comparable results to the cooling constraints derived above. Event rates at neutrino detectors, however, are proportional to the product of the flux and the neutrino interaction cross section. Since low-energy neutrino cross sections scale as $\sigma \sim G_F^2 E_\nu^2$ a ten-fold increase in the neutrino energy can lead to a hundred-fold increase in the event rate.\footnote{More precisely neutrino cross sections scale as $G_F^2 Q_{\rm max}^2$. At low energies $Q_{\rm max}^2 \sim E_\nu^2$, vs. $Q_{\rm max}^2 \sim s \sim 2 m_T E_\nu$ at high energies with $m_T$ the target mass.} This is why bounds from the absence of high-energy events from SN1987A can yield stronger constraints on neutrinophilic bosons compared to standard cooling arguments. 

To compute an exclusion curve from this method, we calculate the flux of neutrinos from the decay $V\rightarrow \nu \bar{\nu}$.
As we discuss below, detection relies on electron neutrinos/anti-neutrinos whose flux arises from oscillation.
We may obtain the differential $\nu_e$ flux, as a function of energy via 
\begin{equation}
    \dv{\Phi_{\nu_e}}{E_\nu} = P(\nu_e|\nu_{\mu,\tau})\frac{{\rm BR}(\nu_{\mu,\tau})}{4\pi R_{\rm 1987A}^2} \int  \dd \omega  \frac{1}{\omega} P(E_\nu|\omega) \frac{\dd L_V}{\dd \omega}~,
\end{equation}
where $P(\nu_e|\nu_{\mu,\tau})$ is the oscillation probability of $\nu_{\mu,\tau}$ into $\nu_e$; $P(E_\nu|\omega)$ is the probability of getting a neutrino with energy $E_\nu$ from a vector with energy $\omega$; ${\rm BR}(\nu_{\mu,\tau})$ is the branching ratio of $V$ into $\nu_\mu$ or $\nu_\tau$ neutrinos; $R_{\rm 1987A}=51.4~{\rm kpc}$ is the distance between Earth and SN1987A and we have expressed the result in terms of the $V$-luminosity function (differential with respect to $\omega$). The expression for $\dd L_V/\dd \omega$ can be inferred from \cref{eq:full_luminosity}, and the factor of $1/\omega$ is necessary to convert luminosity (energy per unit time) of $V$ into an integrated flux.
Going from the rest frame of $V$, where the energy of $\nu_{\mu,\tau}$ is fixed ($E_\nu=m_V/2$), to the star's frame results in a uniform box-distribution for the neutrino energy.   Thus, the function $P(E_\nu|\omega)$ is given by, 
\begin{equation}
    P(E_\nu|\omega) = \frac{\Theta(E_\nu-E_{\nu}^{\rm min})\times \Theta(E_{\nu}^{\rm max}-E_\nu)}{E_{\nu}^{\rm max}-E_{\nu}^{\rm min}} ~,
\end{equation}
where the ends of the box-distribution are determined by the energy of $V$,
\begin{equation}
    E_{\nu}^{(\rm max, \rm min)} = \frac{\omega}{2}(1 \pm \beta )~,
\end{equation}
with $\beta = \sqrt{1-m_V^2/\omega^2}$. 

Neutrinos from a supernova oscillate and at late times appear as incoherent superpositions of mass eigenstates. We therefore multiply the flux of $\nu_\mu$ and $\nu_\tau$ originating at the source by
\begin{equation}
     P(\nu_e|\nu_{\mu,\tau}) = \frac12 \sum_{i=1}^3 \qty[ |U_{\mu i}|^2+  |U_{\tau i}|^2] |U_{e i}|^2 \approx  0.22~,
\end{equation}
irrespective of the neutrino mass ordering or leptonic CP violation~\cite{Esteban:2024eli}.

The detection channel is a combination of inverse beta decay, which only involves $\bar{\nu}_e$ ($\bar{\nu}_e + p \rightarrow n+e^+$), and scattering off oxygen, for both $\nu_e$ and $\bar{\nu}_e$ ($\stackrel{\scriptscriptstyle(-)}{\nu}_e + ^{16}{\rm O} \rightarrow e^\pm + X$).  We compute a total event count, $N_{\rm ev}$, of $e^\pm$ events expected in an energy range $E_{\rm min}\le E_e \le E_{\rm max}$,
\begin{equation}
    N_{\rm ev} = \sum_i \int_0^T \dd t\int_{E_{\rm min}}^{E_{\rm max}} \dd E_e~\epsilon_{\rm det} \int \dd E_\nu ~ \dv{\Phi_{\nu_e}}{E_\nu} \dv{\sigma_{\rm det}^i}{E_e}  N_i ~,
\end{equation}
where $T$ is the exposure time, $\dd \sigma_{\rm det}^i/\dd E_e$ is the detection cross section for target $i$ (that depends on $E_\nu$) differential in the reconstructed energy, $\epsilon_{\rm det}(E_e)$ is a detection efficiency, and $N_i$ is the number of target nuclei for each target type (here H and O) in the fiducial volume of the detector. 

To set constraints we choose to follow a conservative ``rate only'' Poissonian approach; this is a useful compliment to the existing literature, e.g.~\cite{Fiorillo:2022cdq,Akita:2023iwq}, which employ detailed spectral analyses including an assumed SM and BSM spectra, and detector resolution effects. We use the IMB data set \cite{IMB:1988suc} and demand $N_{\rm ev} <3$ in the interval $E_e\in [75~{\rm MeV},~200~{\rm MeV}]$ where IMB saw zero events \cite{Fiorillo:2022cdq}. For the IMB detector we take a fiducial mass of $6.8~{\rm kton}$ of water corresponding to $N_{^{16}\mathrm{O}}=2.3\times 10^{32}$ and $N_{^{1}\mathrm{H}}=4.5\times 10^{32}$, and an exposure time of $T=10~{\rm s}$. We take an energy-independent detection efficiency of $\epsilon_{\rm det}=0.8$ in the energy window of interest. 

For the detection cross section we assume $E_e \geq E_\nu-15~{\rm MeV}$, and therefore integrate from $E_\nu=90~{\rm MeV}$ up to $E_\nu=215~{\rm MeV}$. We do not attempt to model the shape of the differential cross section $\dd \sigma/\dd E_e$ since our single-bin analysis is insensitive to this detail. Again following  Ref.~\cite{Fiorillo:2022cdq}, we take the cross section from Ref.~\cite{Kolbe:2002gk}. This leads to our final formula 
\begin{equation}
    N_{\rm ev}^{({\rm est})} = 0.8\sum_i \int_0^{{\rm 10~s} }\!\!\dd t \int_{90~{\rm MeV}}^{215~{\rm MeV}}\!\! \dd E_\nu ~ \dv{\Phi_{\nu_e}}{E_\nu} \sigma(E_\nu)  N_i ~.
       \label{eq:nu_scattering_event_yield}
\end{equation}
The time integral in \cref{eq:nu_scattering_event_yield} is evaluated numerically using SN profile snapshots to compute the integrand on a discrete time grid. 
We demarcate the boundary between excluded and viable parameter space by  $N_{\rm ev}^{({\rm est})} = 3$ (rounding up to the nearest integer from a 95\% confidence level exclusion assuming Poissonian statistics). The excluded regions are shown in gray in \cref{fig:combined_sn_bounds}. We see that the scattering bound is significantly more powerful than the standard cooling argument. Compared to the more sophisticated analysis carried out in \cite{Akita:2023iwq}, our constraints are somewhat weaker. This can be due to, for example, our inclusion of gravitational redshift (see \cref{sec:sn_appendix}), or their use of full spectral information. 

\subsection{Low-energy supernovae}
\label{sec:low_energy_sn}
It has recently been realized \cite{Caputo:2022mah,Fiorillo:2025sln,Fiorillo:2025yzf} that observations of low-energy supernova place new, and complimentary, constraints on light feebly interacting particles. If a light particle which decays to electromagnetically-interacting final states is produced in the core of a proto-neutron star, then it can efficiently transport energy into the surrounding material of the progenitor star. This can increase the total energy of the explosion, which may be incompatible with observations of relatively dim supernovae.
It is argued in \cite{Caputo:2022mah} that a conservative requirement is that less than one Bethe ($10^{51}~{\rm erg}$) is deposited in the progenitor.  When compared to the typically-employed Raffelt criterion luminosity of $L_R =4.4 \times 10^{52} ~{\rm erg}/{\rm s}$ (emitted over few-second time-scales), this is a smaller amount of energy and therefore offers the possibility of tighter constraints. 

For the case of a $L_{\mu}-L_\tau$ gauge boson, these considerations only become important above the dimuon threshold. 
Below this threshold, in any region in the $L_{\mu}-L_\tau$ parameter space which is compatible with the Raffelt criterion, the $V$'s would deposit at most $10^{47}~{\rm erg}/{\rm s}$ in the progenitor star (assuming all $V$'s decay inside the progenitor), because of the $\sim 10^{-5}$ branching fraction to $e^+e^-$. So this constraint would not be competitive for $m_V < 2m_\mu$. 
Above the dimuon threshold, however, $V\rightarrow \mu^+\mu^-$ has a large branching ratio. 

To set limits on this regime we compute the rate of $V$-emission from the neutrinosphere as in \cref{sec:ballistic}, but now weigh the result by the probability of $V$ decaying inside the progenitor star. We multiply this quantity by the branching ratio ${\rm BR}(V\rightarrow \mu^+ \mu^-)$. Finally, we assume that both muons lose all their kinetic energy in the progenitor (via Bethe-Bloch like energy loss), that the muons decay at rest, and that all energy emitted in the form of neutrinos is lost. We therefore demand that
\begin{equation}
    \bigg\langle P_{\mu^+\mu^-}(E_V) \times \qty(E_V - 2 \langle E_\nu \rangle )\bigg\rangle \leq 10^{51}~{\rm erg}~,
\end{equation}
where $\langle E_\nu \rangle \approx 70~{\rm MeV}$ is the average energy lost to neutrinos per muon decay-at-rest,  $P_{\mu^+\mu^-}(E_V)$ is the probability for $V\rightarrow \mu^+\mu^-$ in the volume of the progenitor star, and the angle brackets denote the integral over the emitted $V$-distribution described in \cref{sec:ballistic}. 
This defines our ``low-energy supernova'' constraint. 
As discussed in \cite{Caputo:2022mah}, although \texttt{s18.8-SFHo-MUONS} is not representative of low-energy supernova, its core is a sufficiently good proxy for the purposes of setting constraints on feebly interacting particles. The resulting bounds are shown by the dotted green lines in \cref{fig:combined_sn_bounds}.

\vspace{-12pt}
\subsection{Results \label{sec:supernovae-results} }
\begin{figure*}[!htbp]
    \centering
    \includegraphics[width=0.85\linewidth]{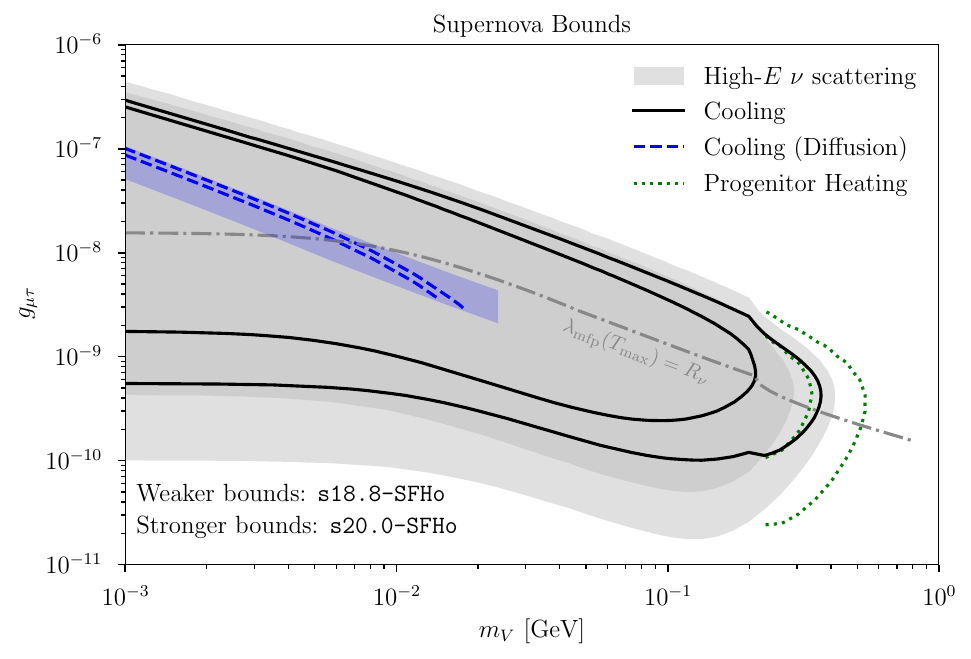}
    \caption{ Supernova bounds on the $L_{\mu}-L_\tau$ gauge boson. We show bounds arising from observed neutrino spectra (gray shading) of SN1987A, cooling constraints computed in the single production/absorption limit (solid black lines) and the diffusive limit (dashed blue lines) and limits on SN progenitor heating (green dotted lines). In each case there are two sets of bounds derived from two SN models, \texttt{s18.8-SFHo-MUONS} and \texttt{s20.0-SFHo-MUONS}. The diffusive cooling curves terminate at couplings where the diffusion approximation breaks down (the region of validity of this approximation is indicated by the light blue shaded region); these curves exclude parameter space below them. The energy deposition bounds exclude parameter space to the left of dotted lines down to $m_V \approx 2m_\mu$ (below this mass, the visible branching fraction of $V$ is negligible). 
    The gray dash-dotted line shows the masses and couplings for which $V$'s emitted from a high-temperature region have a mean free path of order the PNS radius. Above this line we expect $V$ to significantly contribute to energy transport in the SN, and therefore back-react on the SN time-evolution. Constraints derived using fixed SN profiles are therefore uncertain above this line (we thank the anonymous referee for emphasizing this point to us.)
    \label{fig:combined_sn_bounds}}
\end{figure*}
In~\cref{fig:combined_sn_bounds} we combine the results of each of the supernova constraints discussed above. We find $\mathcal{O}(1)$ disagreement with previous calculations in the literature owing to the various improvements we have made (discussed in detail in \cref{sec:sn_appendix}). We have been able to reproduce previous results when replicating the methods used in those papers. Constraints for an extended mass range are presented in \cref{sec:sn_appendix}.

In agreement with~\cite{Fiorillo:2022cdq,Lai:2024mse} we find that the direct detection of high energy neutrinos offers the most powerful probe of $L_{\mu}-L_\tau$ gauge bosons over the vast majority of the parameter space. Unsurprisingly, constraints from (the lack of observation of) high energy neutrinos always surpass the sensitivity of the Raffelt criterion. This is easy to understand: if $L_V\sim L_\nu$ and $\langle E_V \rangle \gg \langle E_\nu \rangle$ and with the neutrino detection cross section being much higher, then the complete absence of high energy events observed in the IMB detector automatically yields a much stronger constraint. 

Interestingly, we find that diffusive transport supplies weaker constraint than both the standard (ballistic) cooling argument and absence of the high-$E_\nu$ signal. This is in contrast to \cite{Cerdeno:2023kqo}, which estimated new reach by a heuristic comparison of diffusive transport time-scales. 
Our approach, instead, makes use of an approximate solution of the Boltzmann equation expanded around the equilibrium solution. This leads to a result which is proportional to temperature gradients, and therefore vanishes (as it must) for a uniform temperature body in thermal equilibrium. Importantly, the approximation used is only valid in a narrow region of parameter space, which is highlighted as a light blue region in \cref{fig:combined_sn_bounds}. This region of validity stems from the inequality in \cref{eq:diffusion_validity}.

Finally, for sufficiently large vector masses, we do indeed find that $V\rightarrow \mu^+\mu^-$ decays inside the progenitor star provide the strongest constraints. They surpass event the ``high $E_\nu$'' probe discussed above. These constraints are computed with a rather conservative limit of $10^{51}~{\rm erg}$ of energy deposition, although they are not as robust as the background-free Poissonian analysis in the case of the high-$E_\nu$ signal. 

We compare this set of constraints against the SHiP sensitivity in~\cref{fig:summary} -- the ``Supernovae'' curves consist of the collection of these four types of bounds for the two progenitor models.
Supernovae considerations provide constraints that are highly complementary to the fixed-target SHiP sensitivity as well as existing constraints from cosmology~\cite{Escudero:2019gzq} and from Borexino~\cite{Altmannshofer:2019zhy,Kelly:2024tvh}.

\section{Conclusions \label{sec:conclusions} }
Two efficient tools for searching for sub-GeV feebly-interacting particles are beam dump experiments with a large decay volume and core-collapse supernovae. The case of a massive vector that couples to the Standard Model's $L_\mu-L_\tau$ current requires special care due to its direct couplings to neutrinos and its highly suppressed couplings to photons, electrons, and hadrons. 

We have studied the phenomenology of a massive $L_\mu-L_\tau$ gauge boson in the context of the upcoming experiment SHiP and in core-collapse supernovae. SHiP is one of the few beam dump experiments with compelling sensitivity to this model. In both cases we have found discrepancies with the literature, with significant numerical impacts for the future sensitivity of SHiP. Our updates to the  supernovae constraint curves for $L_\mu-L_\tau$ reflect both recent improvements in the calculation of the microphysics involved, as well as the inclusion of conceptually new probes of the $L_{\mu}-L_{\tau}$ gauge boson (which substantially extend the reach of the supernovae constraints). 

\begin{acknowledgments}\vspace{-0.8em}
We thank  Thomas Janka, and Daniel Kresse for providing SN profiles, Julien Froustey for discussions about neutrino instabilities, Edoardo Vitagliano, Gustavo Marques Tavares and Shirley Li for discussions on supernovae, Philip Ilten for conversations regarding \textsc{DarkCast}, and Patrick Foldenauer for discussions regarding the details of Ref.~\cite{Bauer:2018onh}. 
RP was supported during most of this project by the Neutrino Theory Network under Award Number DEAC02-07CH11359, by the U.S. Department of Energy, Office of Science, Office of High Energy Physics under Award Number DE-SC0011632, and by the Walter Burke Institute for Theoretical Physics.
Fermilab is managed by FermiForward Discovery Group, LLC, acting under Contract No.89243024CSC000002 with the U.S. Department of Energy, Office of Science, Office of High Energy Physics.  
NB acknowledges support from the Natural Sciences and Engineering Research Council of Canada (NSERC). This research was enabled in part by support provided by Compute Ontario, Calcul Qu{\'e}bec and the Digital Research Alliance of Canada (\href{http://alliancecan.ca}{alliancecan.ca}).
KJK and TZ are supported in part by US DOE Award \#DE-SC0010813.
\end{acknowledgments}

\vfill
\pagebreak

\appendix

\section{Further exploration of SHiP sensitivity}\label{sec:beamdump_appendix}
\begin{figure}
    \centering
    \includegraphics[width=\linewidth]{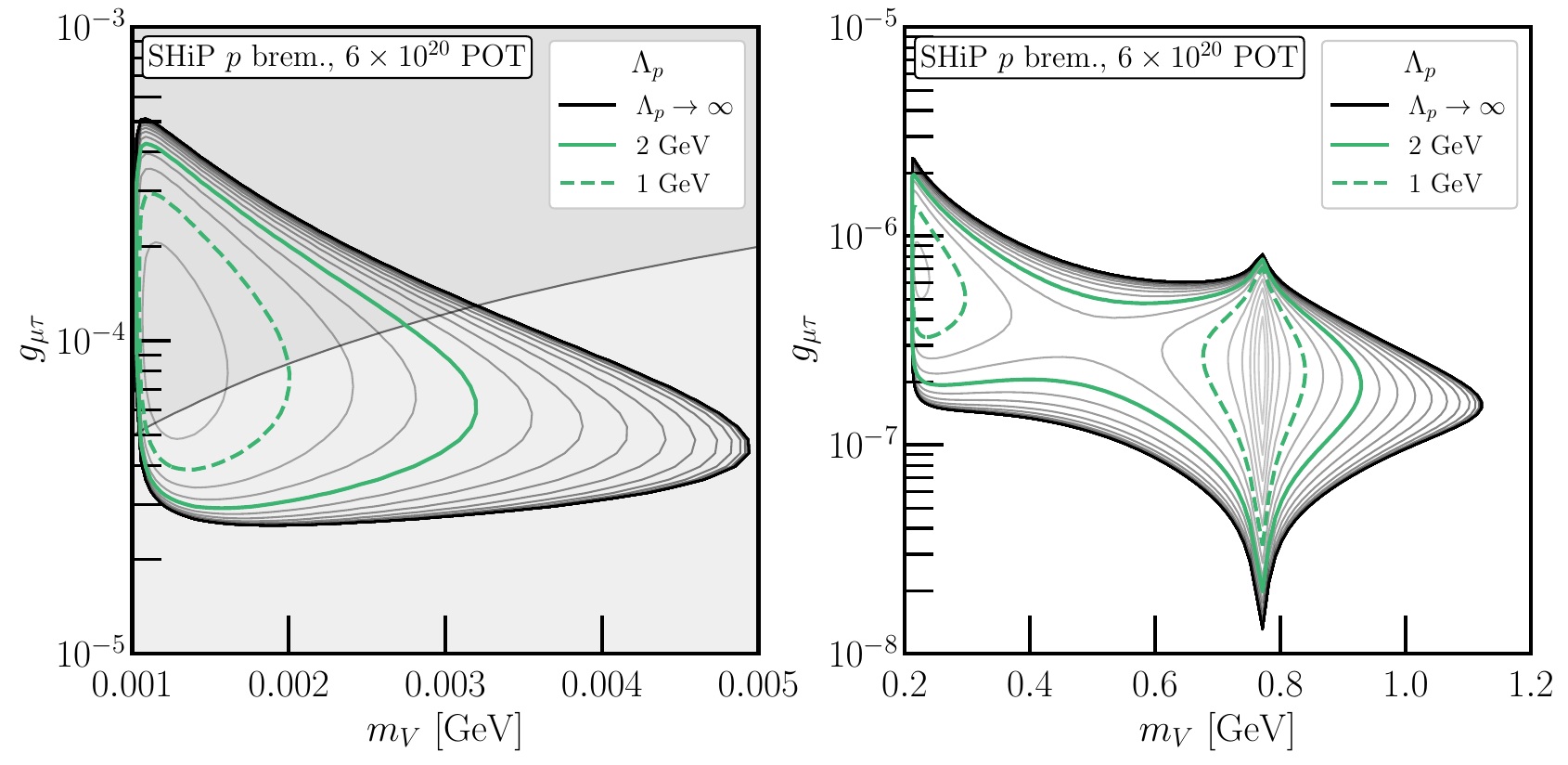}
    \caption{Impact of the parameter $\Lambda_p$ for on-shell protons in proton bremsstrahlung (within the QRA formalism) on SHiP sensitivity. The left (right) panel corresponds to sensitivity below (above) the dimuon threshold. Green lines indicate the sensitivity obtained when taking $\Lambda_p = 1$~GeV (dashed) or $2$~GeV (solid), and the black line corresponds to the $\Lambda_p \to \infty$ limit, where we can effectively ignore this effect. All contours assume $6 \times 10^{20}$~POT exposure.
    \label{fig:QRA_FormFactorImpact}}
\end{figure}

In this appendix, we provide some additional details and further exploration regarding our SHiP beam-dump sensitivity for $L_\mu - L_\tau$ gauge bosons.
First, we provide a comparison against previous literature below the di-muon threshold,\!\footnote{Above the dimuon threshold, comparisons are complicated due to specific choices of how to treat proton bremsstrahlung, which can dominate sensitivity. We focus on this in our discussion surrounding~\cref{fig:QRA_FormFactorImpact}.} where we find substantial differences (our results generally are less powerful than previous expectations) -- we provide some explanations for the sources of discrepancies.
Next, we discuss how, due to the low event-rate expectations under this scenario at SHiP, modest gains in exposure can lead to significant gains in sensitivity across this parameter space.
Finally, we provide more details regarding our simulation of proton bremsstrahlung and the impact of the form-factor parameter $\Lambda_p$ on our results.

\begin{figure}
    \centering
    \includegraphics[width=0.48\linewidth]{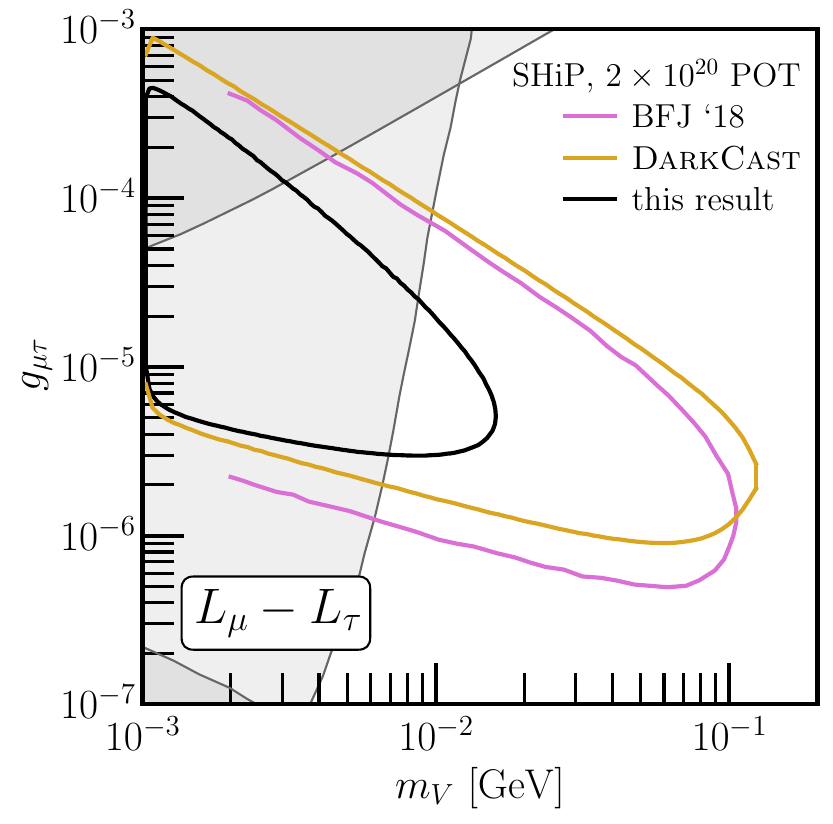}
    \includegraphics[width=0.48\linewidth]{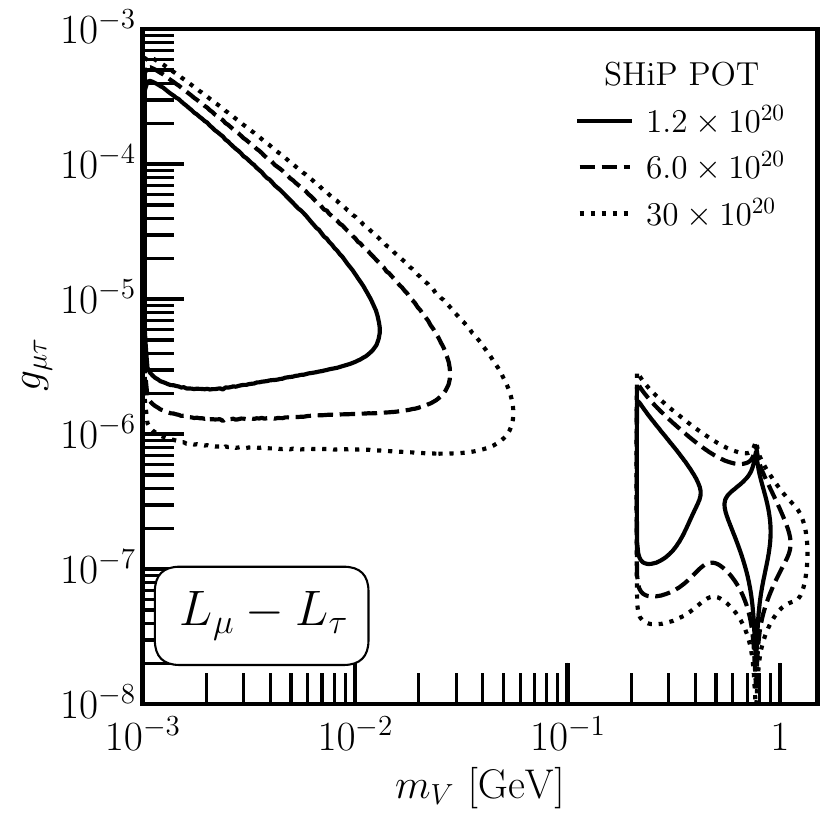}
    \caption{\textbf{Left}: Comparison with previous literature -- for consistency, our results include $V$-production solely from neutral meson decays and proton bremsstrahlung.  BFJ (purple)~\cite{Bauer:2018onh} rescales between models in a different approach, and the \textsc{DarkCast} (gold)~\cite{Ilten:2018crw,Ilten:2022lfq} approach fails when event rates are not substantially large. 
    \textbf{Right}: Illustration of scaling with beam intensity and/or live-time.  
    \label{fig:SHiP_Appendix}
    }
\end{figure}

\paragraph{Comparisons against other approaches}
In~\cref{fig:SHiP_Appendix} (left), we compare our result for SHiP sensitivity (assuming $2.0\times 10^{20}$~POT) against previous literature results -- those of Bauer, Foldenauer, and Jaeckel (BFJ)~\cite{Bauer:2018onh} in purple, and from using \textsc{DarkCast} in orange~\cite{Ilten:2018crw,Baruch:2022esd}. \textbf{Note:} to perform more direct comparisons, in~\cref{fig:SHiP_Appendix} (left), we only include $V$-production from neutral meson decays and from proton bremsstrahlung.

\textbf{Comparison against BFJ:} In Ref.~\cite{Bauer:2018onh}, the production rates from neutral meson decays are calculated using double-differential meson-production cross sections extracted from Ref.~\cite{Aguilar-Benitez:1991hzq}. This reference reports precise measurements of neutral meson production with respect to the mesons' longitudinal momentum, as well as parameterizations capturing the spectra of their transverse momenta. This approach can yield significant differences compared with the the neutral meson spectra we obtain using \texttt{GEANT-4} or \texttt{PYTHIA-8}, which can lead to markedly different sensitivities.
In our case, this leads to sensitivity only reaching ${\sim}20$~MeV as opposed to ${\sim}100$~MeV from Ref.~\cite{Bauer:2018onh}, indicated by the purple line in~\cref{fig:SHiP_Appendix}(left).
Note that our sensitivity here only includes neutral meson decay and proton bremsstrahlung; in~\cref{fig:SHiP_Sensitivity}, the electron/positron bremsstrahlung and charged-meson decays provide additional sensitivity for small masses and small couplings.

\textbf{\textsc{DarkCast} Comparison:} We construct the orange line in~\cref{fig:SHiP_Appendix} (left) by taking our own dark-photon sensitivity (using only neutral-meson decays and proton bremsstrahlung) and recasting with \textsc{DarkCast} to the $L_\mu - L_\tau$ gauge boson scenario. 
Our simulations regarding dark-photon sensitivity are discussed in detail in Ref.~\cite{Zhou:2024aeu}. When making comparable assumptions regarding detector geometry, exposure, and simulation tools (e.g.~\texttt{PYTHIA-8} for meson production), we find excellent agreement with official projections~\cite{SHiP:2020vbd}.

We find that \textsc{DarkCast} recasts the bottom edge of sensitivity curves by assuming that the sensitivity is attained in the long-lived-particle limit, i.e.~that event rates scale like $g_{\mu\tau}^4$.
Additionally, the upper edge of sensitivity curves is determined by using arguments of lab-frame particle lifetimes in the different models.
If we consider the event-rate expectation as a function of $g_{\mu\tau}$ for a fixed $m_V$ (for instance $m_V \approx 10$~MeV), our analysis finds that the maximum event rate occurs for $g_{\mu\tau} \approx 10^{-5}$ and it \textit{barely} exceeds 10 events, which we use as a benchmark for sensitivity.
In such situations, where the maximum expected event rate fails to reach significantly large values, the recasting procedure of \textsc{DarkCast} fails -- one can interpret this as the asymptotic limit (event rates $\propto g_{\mu\tau}^4$) and the decay-length-limited region ($V$ fails to reach the decay chamber) coming too close to one another.
We have verified that if we scale up our expected SHiP exposure by a large, arbitrary factor (e.g.~$10^5$), our sensitivity and the curves projected by \textsc{DarkCast} agree.

\paragraph{Scaling with SHiP Exposure}
The nominal expected exposure for the SHiP experiment is $6.0\times 10^{20}$~POT~\cite{Albanese:2878604}, however projections have often used the older nominal value of $2.0\times10^{20}$~POT.
Our $L_\mu - L_\tau$ sensitivity has demonstrated that, for either of these exposure choices, the most events expected across parameter space barely exceeds the ten events we require as a sensitivity threshold.
This means that the gains in sensitivity for modest increases in exposure scale more aggressively than when in the large-exposure limit.
We demonstrate this in~\cref{fig:SHiP_Appendix} (right), where three different POT scalings ($1.2\times 10^{20}$, solid; $6.0\times 10^{20}$, dashed; $30\times 10^{20}$, dotted) are chosen.
A five-fold increase in exposure can improve the reach in terms of $g_{\mu\tau}$ by factors of a few, and the reach in terms of mass by factors of two or so -- greater gains than in models in which the expected event rates are significantly larger (e.g., for dark photons).

\paragraph{Proton bremsstrahlung details}
In~\cref{sec:beam-dumps}, we discussed the simulation of proton bremsstrahlung of $V$, utilizing the formalism detailed in Ref.~\cite{Foroughi-Abari:2021zbm,Foroughi-Abari:2024xlj} of the quasi-real approximation.
The largest uncertainty in the production rate from proton bremsstrahlung comes from the form factor regarding whether the intermediate proton in this emission is on-shell. 
In our resulting SHiP sensitivity in~\cref{fig:SHiP_Sensitivity}, we assumed $\Lambda_p \to \infty$ to neglect this effect; the authors of Ref.~\cite{Foroughi-Abari:2024xlj} advocate for a choice $1$~GeV $\lesssim \Lambda_p \lesssim 2$~GeV to better agree with experimental data of SM $\rho^0$ production. 
We showed in~\cref{fig:flux10MeV} the impact of varying $\Lambda_p$ from $1$~GeV to $\infty$, leading to an order of magnitude or so reduction in the flux at high energies when $m_V = 10$~MeV.
At heavier $m_V$, this impact is even larger, suppressing fluxes (relative to the $\Lambda_p \to \infty$ limit) by two orders of magnitude or more.

We show the sensitivity using proton bremsstrahlung only and an assumed exposure of $6 \times 10^{20}$~POT at SHiP in~\cref{fig:QRA_FormFactorImpact} -- the two panels are separated only to identify the split regions of parameter space below (left) and above (right) the di-muon threshold where we have sensitivity.
The dark black line corresponds to $\Lambda_p \to \infty$, matching the proton-bremsstrahlung sensitivity we present in~\cref{fig:SHiP_Sensitivity}. In each panel, the green lines correspond to the choices motivated by Ref.~\cite{Foroughi-Abari:2024xlj} -- $1$~GeV (dashed) and $2$~GeV (solid).
The spacing between subsequent contours corresponds to factors of ${\sim}1.25$ scaling of $\Lambda_p$.

Overall, we find that the specific choice of $\Lambda_p$ has a dramatic impact on the resulting experimental sensitivity, especially for $m_V > 2 m_\mu$ (right panel of~\cref{fig:QRA_FormFactorImpact}). We hope that this spurs further theoretical advancement of the QRA formalism and detailed experimental analyses to validate this approach.

\section{Emission of \texorpdfstring{$L_\mu-L_\tau$}{Lmu-Ltau} Bosons From Supernova\label{sec:sn_appendix}}
We estimate the luminosity due to ballistic energy loss via the $L_\mu-L_\tau$ gauge boson using 
\beq
L_V = \int_0^{R_\text{max}} dr 4\pi r^2 N(r)^2 \int_{m_V/N(r)}^{\infty} d\omega \frac{k \omega}{2 \pi^2} \omega ~\Gamma_{\text{prod}}(r,\omega)~O(r, \omega)~,  
\label{eq:full_luminosity}
\eeq
where $N(r)$ is the gravitational lapse and:
\beq
 O(r, \omega)  = \langle \e^{-\tau}\rangle (r, \omega)  = \frac{1}{2} \int_{-1}^{+1} d c_\theta \exp \left[-\int_0^{s_\text{max}} d s\frac{ \Gamma_{\text{abs}}^*(\sqrt{r^2+s^2+2 r s c_\theta},\omega)}{v_V}\right]~,
\label{eq:angular_averaged_absorption_factor}
\eeq
%
is the $V$-direction-averaged opacity factor which encodes the probability of $V$ to escape from its production point at radius to beyond $R_{\mathrm{max}}$ without interacting or decaying~\cite{Caputo:2022rca}. The argument of $\Gamma_{\rm abs}^*$ given by $r'(s)=\sqrt{r^2+s^2+2 r s c_\theta}$ is the radial position along the path length with $s$ parameterizing the distance from $r$ to the edge of the sphere at $R_{\rm max}$. The distance from the production point to $R_{\mathrm{max}}$ along a chord determined by emission angle $c_\theta= \cos\theta$ is
\beq
s_\text{max} = -c_\theta r + \sqrt{R_\text{max}^2 - r^2 + r^2 c_\theta^2}~;
\eeq
the reduced absorption rate
\beq
\Gamma_{\text{abs}}^* = \left(1-e^{-\omega/T}\right) \Gamma_{\text{abs}}~,
\label{eq:reduced_absorption_rate}
\eeq
encodes both absorption and the reduction of absorption due to spontaneous emission; the speed of the vector is
\beq
v_V = \sqrt{1 - m_V^2/\omega^2}~.
\eeq

\Cref{eq:full_luminosity} contains several notable differences compared to some previous works focusing on $L_\mu-L_\tau$:
\begin{itemize}
    \item Full angular averaging. The pathlength taken by $V$ through the SN depends on the direction in which it was emitted. Many estimates have assumed outward radial paths (i.e., fixed $c_\theta = 1$), which minimizes the probability of absorption. As emphasized in \cite{Caputo:2022rca,Lai:2024mse}, this approximation can change the $V$-luminosity by a factor of a few. 
    \item Vector speed $v_V$. The optical depth factor in \cref{eq:angular_averaged_absorption_factor} must contain a factor $1/v_V$, since slower particles spend more time in the SN before escaping. In particular, a particle at rest never escapes. This factor was absent in many dark sector studies; it was re-introduced in \cite{Caputo:2022rca}.
    \item Reduced absorption rate, \cref{eq:reduced_absorption_rate}, encodes the fact that stimulated emission can occur along the direction of propagation of $V$. This emission effectively reduces the probability of absorption. This fact can be seen from the Boltzmann equation along a given dark vector's ray~\cite{Caputo:2022rca}. In the context of $L_\mu - L_\tau$ this modification was introduced in \cite{Lai:2024mse}, following~\cite{Caputo:2022rca}.
    \item The maximum production radius $R_{\mathrm{max}}$ is the same as the minimum radius beyond which the $V$ has to decay to constitute energy loss. Here we identify $R_{\mathrm{max}}$ with the neutrinosphere radius $R_\nu \approx 25$ km, as advocated in \cite{Lai:2024mse}. This choice is motivated by the exact solution to the Boltzmann equation from which \cref{eq:full_luminosity} is derived~\cite{Caputo:2022rca}. Compared to \cite{Croon:2020lrf}, who set the minimum absorption radius to the neutrino gain radius, $\sim 100$ km, our choice  improves the reach at small couplings/short $V$-lifetimes. 
    \item Inclusion of gravitational effects. As discussed in Appendix C of \cite{Caputo:2022mah}, the gravitational well of the PNS leads to time-dilation of the production rate, and a red-shifting of the energy of the emitted particles.  As a result, the differential luminosity as seen by a distant observer is suppressed by $N(r)^2$. In the numerical models used here, the lapse can be as small as $\sim 0.65$ in the inner regions of the PNS, leading to a significant reduction of the luminosity. In addition to these effects, the large escape velocity modifies the lower bound of the $\omega$ integral to be above $m_V$: $\omega_{\mathrm{min}} = m_V/N(r)$. In a Schwarzschild spacetime $\omega_{\mathrm{min}} \approx m_V (1 + GM/r)$, which is equivalent to the more familiar condition $v > v_{\mathrm{esc}} = \sqrt{2 GM/r}$ in the non-relativistic limit~\cite{Magill:2018jla}. 
\end{itemize}

The main production channels for $V$ are: 
\begin{itemize}
    \item Inverse neutrino decay. The production rate the for this channel is computed via 
    \beq
    \begin{split}
\Gamma_{\bar{\nu}\nu\to V}^\mathrm{prod}(\omega) & =\frac{1}{2 \omega} \int \frac{d^3 p_{1}}{2 E_{1}(2 \pi)^3} \frac{d^3 p_{2}}{2 E_{2}(2 \pi)^3}(2 \pi)^4\delta^4\left(p_{1}+p_{2}-k_{V}\right)        \\
&\times \sum_{\text{pols}}\left|\mathcal{M}_{V\rightarrow \bar{\nu} \nu}\right|^2 f_{\bar\nu}(p_1;-\mu_\nu) f_\nu(p_2; \mu_\nu) ~,
    \end{split}
\label{eq:production_rate_inv_decay}
\eeq
where $\sum_{\text{pols}}$ indicates a sum over initial and final spin states. 
Note that the Bose factor $(1 + f_V(\omega))$ that would be present on the right hand side is instead accounted for by the reduction of the absorption rate, i.e., \cref{eq:reduced_absorption_rate}~\cite{Caputo:2022rca}. Neutrinos are highly degenerate inside the PNS, with $\mu_\nu \sim 100\;\MeV$. We find that contrary to \cite{Croon:2020lrf,Lai:2024mse}, the chemical potential dependence in $\nu\bar{\nu} \to V$ does not cancel (this simplification is valid only the Maxwell-Boltzmann limit, $\omega/T \gg 1$). The result of performing the phase space integral is 
\beq
\Gamma_{\bar{\nu}\nu\to V}^\mathrm{prod}(\omega) = 
 \frac{m_V}{\omega}3\Gamma_{V\rightarrow \bar{\nu} \nu}  I( \mu_{\nu}/T,\omega/T, v_V)~,
\label{eq:production_rate_inv_decay_simplified}
\eeq
where 
\beq
\Gamma_{V\rightarrow \bar{\nu} \nu} = \frac{1}{6}\alpha_V m_V~,
\eeq
is the vacuum decay rate into one flavor of left-handed neutrinos, and 
\beq
I(\alpha,\beta, v) = \frac{\log \left(\frac{\left(e^{\alpha }+e^{\frac{1}{2} \beta  (v+1)}\right) \left(e^{\alpha +\frac{1}{2} \beta  (v+1)}+1\right)}{e^{\alpha +\beta
   }+\left(e^{2 \alpha }+1\right) e^{\frac{1}{2} \beta  (v+1)}+e^{\alpha +\beta  v}}\right)}{\left(e^{\beta }-1\right) \beta  v}~.
   \label{eq:inverse_decay_integrand_correct}
\eeq
Note that the definition of vacuum decay rate includes an average over the initial polarizations of $V$, whereas the thermal production rate, \cref{eq:production_rate_inv_decay}, is instead a sum. This difference is accounted for by the explicit factor of $3$ in \cref{eq:production_rate_inv_decay_simplified}, as pointed out in \cite{Lai:2024mse}.

In the Maxwell-Boltzmann limit ($\beta\gg 1$ and $v \ll 1$) and neglecting chemical potentials $(\alpha = 0)$ we find that $I(\alpha,\beta,v) \sim e^{-\omega/T}$ and the result matches \cite{Lai:2024mse}. Another non-trivial check of \cref{eq:inverse_decay_integrand_correct} is a direct calculation of  $\Gamma_{V\to \bar{\nu}\nu}^\mathrm{abs}(\omega)$, which is determined by an integral similar to \cref{eq:production_rate_inv_decay} but with $f_{\bar\nu} f_\nu\to (1-f_{\bar\nu})(1 - f_\nu)$. In this calculation the analogue of $I(\alpha,\beta,v)$ is such that it ensures the detailed balance condition 
\beq
\Gamma_{\bar{\nu}\nu\to V}^\mathrm{prod} = e^{-\omega/T} \Gamma_{V\to \bar{\nu}\nu}^\mathrm{abs}~,
\eeq
is satisfied, in agreement with general arguments of \cite{Weldon:1983jn}. Finally, replacing the Fermi-Dirac distributions by Bose-Einstein reproduces the diphoton results of \cite{Caputo:2022mah}.

In the SN profiles we use for our numerical results, $\nu_{\mu}$ and $\nu_\tau$ have different chemical potentials, so we apply these formulae to the two flavours separately. Chemical potentials for $\mu$ and $\nu_\mu$ arise from $\beta$ equilibrium with $e$ and $\nu_e$, and from differences in diffusion of neutrinos compared to antineutrinos out of the PNS~\cite{Bollig:2017lki}. Tau neutrino chemical potential can only arise from the latter effect since $\tau^\pm$ are too heavy to be produced; however this effect is relatively small and the simulations we use do not present $\mu_{\nu_\tau}$, so we set $\mu_{\nu_\tau} = 0$.
\item Inverse muon decay. This calculation is nearly identical to $\bar{\nu}\nu\to V$ above with the result 
\beq
\Gamma_{\bar{\mu}\mu\to V}^\mathrm{prod}(\omega) = 
 \frac{m_V}{\omega}3\Gamma_{V\rightarrow \bar{\mu} \mu}  I\left( \mu_\mu/T,\omega/T, v_V\sqrt{1 -4m_\mu^2/m_V^2}\right)~.
\label{eq:production_rate_inv_decay_muons}
\eeq
where the vacuum decay rate of $V$ to both muon chiralities is 
\beq
\Gamma_{V\rightarrow \bar{\mu} \mu} = \frac{\alpha_{V} m_{V}}{3} \left(1 + \frac{2m_\mu^2}{m_{V}^2}\right)\sqrt{1 - \frac{4m_\mu^2}{m_{V}^2}}~.
\eeq
\item Compton scattering. Production via $\mu  \gamma \to \mu V$ is computed by 
\beq
\begin{split}
\Gamma_{\mu \gamma \to \mu V}^\mathrm{prod}(\omega)& =\frac{1}{2 \omega} \int \frac{d^3 p_{1}}{2 E_{1}(2 \pi)^3}  \frac{d^3 p_{2}}{2 E_{2}(2 \pi)^3}
\frac{d^3 k_\gamma}{2 E_{\gamma}(2 \pi)^3}(2 \pi)^4\delta^4\left(p_{1}+k_\gamma
-p_2-k_{V}\right)\\
&\times \sum_{\text{pols}}\left|\mathcal{M}\right|^2 f_{\mu}(E_1;\mu_\mu) f_\gamma(E_\gamma)(1-f_\mu(E_2;\mu_\mu)).
\label{eq:production_rate_compton}
\end{split}
\eeq
This rate is usually evaluated in the non-relativistic muon limit, and neglecting their recoil; energy conservation then enforces $E_\gamma\approx \omega$. It is also conventional to write this rate in terms of the cross-section:
\beq
\sigma ( \mu \gamma \to \mu V) \approx 
\frac{8\pi \alpha\alpha_V}{3m_\mu^2} \left(1 + \frac{m_V^2}{2\omega^2}\right) \sqrt{1 - \frac{m_V^2}{\omega^2}}~.
\label{eq:compton_xsec}
\eeq
Terms of $\mathcal{O}(m_V^2/\omega^2)$ in the parenthesis are typically dropped since this process only becomes relevant for $m_V \lesssim \;\MeV$, well below the characteristic energy scales of the SN (see footnote 4 in \cite{Croon:2020lrf}).
Note that the phase space integrals in \cref{eq:production_rate_compton} are appropriate for $\sigma(\mu V\to \mu \gamma)$, not $\sigma(\mu \gamma\to \mu V)$; this leads to  additional factors of $v_V$ that have been missed in some works. Direct evaluation of \cref{eq:production_rate_compton} gives
\begin{align}
\Gamma_{\mu \gamma \to \mu V}^\mathrm{prod}(\omega)
& =  n_\mu F_{\mathrm{deg}} \frac{1}{e^{\omega/T} - 1} \times 3 v_V \sigma(\mu V\to \mu \gamma)~,\\
& = \frac{1}{v_V} n_\mu F_{\mathrm{deg}} \frac{2}{e^{\omega/T} - 1} \times \sigma(\mu \gamma\to \mu V)~,
\label{eq:production_rate_compton_simplified}
\end{align}
where $n_\mu$ is the muon number density (including the chemical potential) and 
 \beq
F_{\mathrm{deg}} = \frac{1}{n_\mu} \int \frac{2 d^3 p}{(2 \pi)^3} f_\mu(E,\mu_\mu)\left(1-f_\mu(E,\mu_\mu)\right)~,
\eeq 
encodes the effects of Pauli blocking~\cite{Raffelt:1996wa,Bollig:2020xdr}. We used 
\beq
\frac{\sigma(\mu \gamma\to \mu V)}{\sigma(\mu V\to \mu \gamma)} = \frac{3}{2} v_V^2~,
\eeq 
to get the second line in \cref{eq:production_rate_compton_simplified}. \cref{eq:production_rate_compton_simplified} matches other published results up to the extra factors of $1/v_V$~\cite{Croon:2020lrf,Lai:2024mse}, and $2$. The velocity factor arises because the $V$ and $\gamma$ phase space volumes differ when $E_\gamma = \omega$ due to $m_V\neq 0$. The factor $2$ is the number of photon spin states, which is needed to cancel the averaging present in the definition of the cross-section, but absent in the definition of the thermal rate, \cref{eq:production_rate_compton}.

The absorption rate is determined by detailed balance
\beq
\Gamma_{\mu \gamma \to \mu V}^\mathrm{abs}(\omega) = e^{\omega/T}\Gamma_{\mu \gamma \to \mu V}^\mathrm{prod}(\omega)~.
\eeq
\item Bremsstrahlung-type reactions $\mu p \to \mu p +V$~\cite{Croon:2020lrf} and $p n \to p n + V$~\cite{Lai:2024mse} have been estimated previously and found to be negligible in the parameter space of interest to us. We therefore do not include these processes in our calculations. 
\end{itemize}

\begin{figure}
    \centering
    \includegraphics[width=0.8\textwidth]{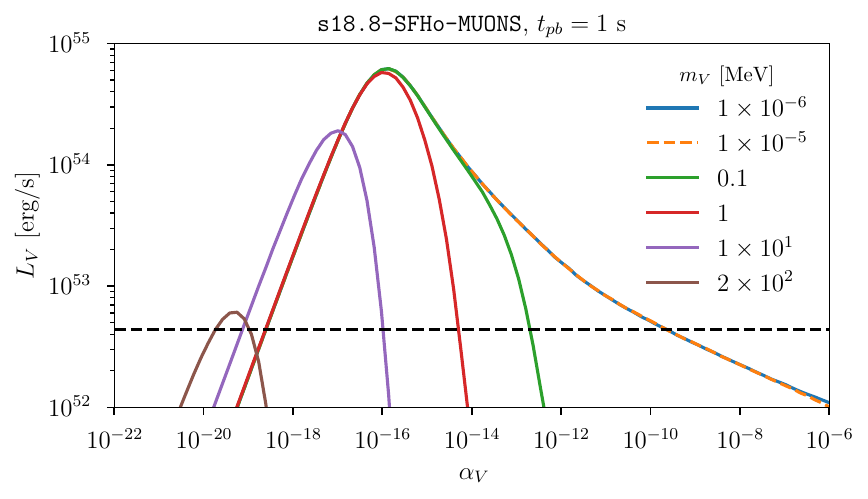}
    \caption{Supernova $V$ luminosity as a function of $\alpha_V$ for several masses, evaluated using the \texttt{s18.8-SFHo-MUONS} model at 1 second post-bounce. The luminosity used to estimate the Raffelt cooling bound is shown as the black dashed line.}
    \label{fig:luminosity_snapshot}
\end{figure}

In \cref{fig:luminosity_snapshot} we show the $V$ luminosity at 1 second post-bounce, evaluated using \cref{eq:full_luminosity} and the physics inputs described above. Low couplings correspond to the long-lifetime regime, where the vectors decay outside of the PNS, while larger couplings correspond to the trapping regime. The main processes responsible for trapping are mass-dependent: above the MeV scale, $V\to \nu\nu$ dominates; at lower masses, absorption is mainly through the Compton process, whose rate is mass-independent (see \cref{eq:compton_xsec}) as $m_V\to 0$.

\begin{figure}
    \centering
    \includegraphics[width=0.8\textwidth]{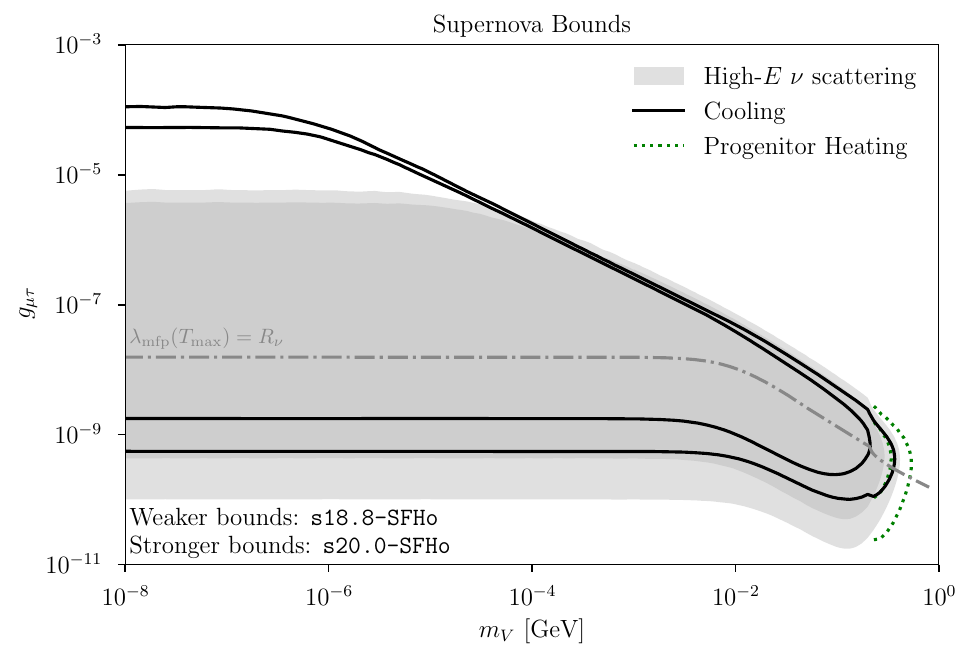}
    \caption{Supernova bounds on the $L_{\mu}-L_\tau$ gauge boson, extended to lower masses compared to \cref{fig:combined_sn_bounds}.}
    \label{fig:supernova_bounds_extended}
\end{figure}

We collect several SN1987A constraints in \cref{fig:supernova_bounds_extended}. This figure is similar to \cref{fig:combined_sn_bounds}, but extended to lower masses; it can be compared directly to the results of \cite{Croon:2020lrf,Lai:2024mse}. Compared to these works, our cooling bounds are slightly weaker due to the computational differences described above. One qualitative difference with \cite{Lai:2024mse} at low masses is the presence of a ``plateau'' in their luminosity calculation. This corresponds to the limit where $V$ interactions are so strong that emission takes place only from a thin shell of the PNS. In this limit, the luminosity given by \cref{eq:full_luminosity} is independent of $\alpha_Z$, as argued in \cite{Lai:2024mse}. While we do observe such effects in our numerical calculations, they occur at smaller luminosities (e.g., below the range shown in \cref{fig:luminosity_snapshot}) and do not affect our exclusion contours.

Finally, we note two important features of the derived constraints. First, the parameter space with $m_V\lesssim 10\;\MeV$ is excluded by $\Delta N_{\mathrm{eff}}$ bounds from BBN and CMB in standard cosmologies. Second, much of the constrained region at larger couplings (above the dash-dotted lines in \cref{fig:combined_sn_bounds,fig:supernova_bounds_extended}) is subject to significant theoretical uncertainties: in this parameter space $V$ is expected to modify the dynamics of the supernova, so using a fixed profile that includes only SM physics is likely incorrect.


\section{Supernova Gamma Ray Bounds}
\label{sec:sn_gamma_ray}
In this Appendix we investigate whether $\gamma$ ray observations can constrain the $L_\mu-L_\tau$ parameter space. In axion-like particle and dark photon models it was found that $\gamma$ ray emission, either from direct (e.g., $a\to \gamma\gamma$~\cite{Caputo:2021rux}) or radiative (e.g., $A'\to e^+ e^- \gamma$~\cite{DeRocco:2019njg,Caputo:2025avc}) decays provide some of the strongest bounds for light mediators with long lifetimes. These bounds come from Solar Maximum Mission (SMM) observations at around the time of the SN1987A event, and from the diffuse $\gamma$ ray emission from all previous supernovae. In the $L_\mu-L_\tau$ model, photons are produced in radiative decays $V\to e^+e^- \gamma$ and $V\to \mu^+\mu^- \gamma$. 
Note that in addition to the $\sim \alpha/\pi$ radiative branching fraction, $V\to e^+e^-$ is also very suppressed by the induced kinetic mixing.
The dimuon channel instead features an $\mathcal{O}(1)$ branching, but it is only open near the kinematic threshold for producing particles in the SN; moreover the gamma ray observables require smaller coupling for the $V$ to exit the progenitor compared to cooling, neutrino scattering and low energy supernova bounds.
 Despite these considerations, it is not immediately clear that the gamma ray bounds are irrelevant as they are typically much stronger than standard cooling arguments. We will find that unlike the axion (with tree-level direct decay to photons) or the dark photon (with unsuppressed decay to $e^+e^-$), $\gamma$ ray observations do not provide a new constraint on $L_\mu-L_\tau$. 

\subsection{$\gamma$ Rays from SN1987A}
SMM measurements place an upper limit on $\gamma$ ray fluence in a time window around the arrival of SN1987A neutrinos; these limits are summarized in Table II of \cite{Caputo:2021rux}. We estimate the fluence following~\cite{Oberauer:1993yr,Jaffe:1995sw}. Our starting point is the fluence differential in lab frame photon energy $k$ and time of their arrival $t$ (with $t=0$ corresponding to the neutrino signal arrival):
\begin{equation}
  \frac{dN}{dk\,dt} = \frac{\br(e)}{4\pi R_{\rm 1987A}^2\,\tau_V}
  \int d\omega\; \frac{dN_V}{d\omega}
  \int_{-1}^{+1} d\bar\mu\;
  f\!\left[\frac{k}{\gamma(1+v\bar\mu)},\;\bar\mu\right]
  e^{-\gamma(1+v\bar\mu)\,t/\tau_V}
  \theta(v t_d - R_\mathrm{Prog})
  \label{eq:JT15}
\end{equation}
This quantity has SI dimensions of $\mathrm{cm}^{-2}\text{MeV}^{-1}\text{s}^{-1}$. Here $\br(e) \sim 10^{-5}$ is the branching fraction of $V \to e^+e^-$; $R_{\rm 1987A}$ is the distance to SN1987A; $dN_V/d\omega$ is the time-integrated number spectrum of $V$'s emitted from the SN (in \cite{Jaffe:1995sw} this is called $L_\#(\omega)$); $\bar{\mu}$ is the cosine of $\gamma$ emission angle \emph{in the rest frame of $V$}; $f(\bar{k},\bar{\mu})$ is the $V$-rest-frame spectrum of $\gamma$ (differential in rest-frame energy $\bar{k}$ and $\bar{\mu}$); the exponential factor encodes probability of $V$ to survive until decay time $t_d = \gamma^2 (1+\bar{\mu}v)t$ (which leads to the photon emitted in the rest-frame direction $\bar{\mu}$ to be detected at time $t$); and the step function enforces the decays to be outside of the progenitor (where the photon would be absorbed). 
This equation is general up to the assumptions that the $V$ decays near the SN (i.e., the decay time $t_d$ is $\ll R_{\rm 1987A}$), and that $V$'s exit the progenitor on radial tragectories (a good assumption since $R_{\mathrm{PNS}} \ll R_{\mathrm{Prog}}$). 

First, let us relate the number spectrum of emitted $V$'s to the luminosity evaluated previously. Dividing \cref{eq:full_luminosity} by $\omega$ and integrating over post-bounce time $t_{pb}$ gives:
\beq
\frac{dN_V}{d \omega} = \frac{1}{\omega}\int dt_{pb} L_V(\omega, t_{pb}).
\label{eq:V_number_spectrum}
\eeq
As for the neutrino scattering signal (\cref{sec:high_E_nu}) we evaluate this integral numerically using SN profile snapshots. The dimensions of this quantity are $1/\MeV$.

Next let us make several assumptions to simplify the form of~\cref{eq:JT15}. The $\gamma$ spectrum in $V\to e^+ e^- \gamma$ is in general a complicated function of photon energy and angle (see the appendices in \cite{DeRocco:2019njg} for  exact results). Following \cite{Jaffe:1995sw} we make two approximations for $f$. First, we assume that it is independent of angle: $f(\bar{k},\bar{\mu}) = f(\bar{k})/2$. Second, we will make the soft-photon approximation for $f(\bar{k})$:
\beq
f(\bar{k}) = \frac{\alpha}{\pi} \frac{1}{\bar{k}}.
\label{eq:approx_photon_spectrum}
\eeq

The total fluence is obtained by integrating \cref{eq:JT15} over the observation time $t\in [0,T_{\mathrm{obs}}]$ and energy $k\in [k_\mathrm{min}, k_\mathrm{max}]$ windows. The approximations above enable the $\bar{\mu}$, $t$ and $k$ integrals to be performed analytically; the $\omega$ integral (as well as the integrations in the definition of $L_V$) must be done numerically. After the analytic steps we find to total fluence to be 
\beq
N = \frac{\br(e)}{4\pi R_{\rm 1987A}^2}\frac{\alpha}{2\pi}\int d\omega\; \frac{dN_V}{d\omega} \ln \left[\frac{\min(\omega, k_\mathrm{max})}{k_\mathrm{min}}\right]\mathcal{I}(\omega)
\label{eq:fluence_full}
\eeq
where $\mathcal{I}$ is the result of carrying out the $\bar{\mu}$ and $t$ integrals under the approximations outlined above:
\beq
\mathcal{I} = 2 e^{-B} - \frac{1}{v A}\left[e^{-A(1-v)} - e^{-A(1+v)}\right],\quad A =  \frac{\gamma T_{\mathrm{obs}}}{\tau_V},\; B = \frac{R_{\mathrm{Prog}}}{v\gamma \tau_V}
\eeq
Note that $R_{\mathrm{Prog}} = 100$ s, while for SMM $T_{\mathrm{obs}} = 223$ s, so typically $B \ll A$ (if $\gamma \gg 1$). In the long lifetime limit, $A\to 0$ and $\mathcal{I} \approx 2 A$ the fluence reduces to 
\beq
N|_{A\to 0} = \frac{\br(e) T_{\mathrm{obs}}}{4\pi R_{\rm 1987A}^2 m_V \tau_V}\frac{\alpha}{\pi} \int d\omega L_V \ln \left[\frac{\min(\omega, k_\mathrm{max})}{k_\mathrm{min}}\right]
\eeq
which reproduces the long-lifetime result in \cite{Jaffe:1995sw} (given there in differential form). This limit is valid when $A\ll 1$ which translates to couplings of 
\beq
\alpha_V < 10^{-25} \left(\frac{30\;\MeV}{T_{SN}}\right)\left(\frac{223\text{ s}}{T_{\mathrm{obs}}}\right)
\eeq
where we took $\omega \sim 3 T_{SN}$. This simple result can be used to make an order of magnitude estimate for the fluence in by noting that the $\omega$ integral essentially gives, $E_V$, the total energy emitted in $V$'s; for $m_V = 10\;\MeV$ we obtain roughly $E_V \sim 10^{50}\;\mathrm{erg} (\alpha_V / 10^{-22})$ (this estimate does not include the mass dependence of this quantity). This gives (setting the log to $\mathcal{O}(1)$)
\beq
N|_{A\to 0} \sim 10^{-4}\;\mathrm{cm}^{-2} \times \left(\frac{\alpha_V}{10^{-25}}\right)^2
\eeq

Another simple limit is a short-lifetime limit $A\gg 1$ with $B\ll 1$, valid for 
\beq
10^{-25} \left(\frac{30\;\MeV}{T_{SN}}\right)\left(\frac{223\text{ s}}{T_{\mathrm{obs}}}\right) \ll \alpha_V \ll 
2\times 10^{-23} \left(\frac{T_{SN}}{30\;\MeV}\right)\left(\frac{10\;\MeV}{m_V}\right)^2
\eeq
Here $\mathcal{I} \approx 2$ and the fluence simplifies to 
\beq
N|_{A\gg 1, \; B\ll 1 } = \frac{\br(e)}{4\pi R_{\rm 1987A}^2}\frac{\alpha}{\pi}\int d\omega\; \frac{dN_V}{d\omega} \ln \left[\frac{\min(\omega, k_\mathrm{max})}{k_\mathrm{min}}\right].
\eeq
We can also use this expression for an order-of-magnitude limit; now the $\omega$ integral is roughly $E_V/\langle \omega \rangle$, so that 
\beq
N_{A\gg 1, \; B\ll 1 } \sim 10^{-2}\;\mathrm{cm}^{-2} \times \left(\frac{\alpha_V}{10^{-23}}\right).
\eeq
For larger couplings $B\gtrsim 1$ and $V$ decays inside the progenitor; the fluence is exponentially suppressed in this regime.

\begin{figure}
    \centering
    \includegraphics[width=\textwidth]{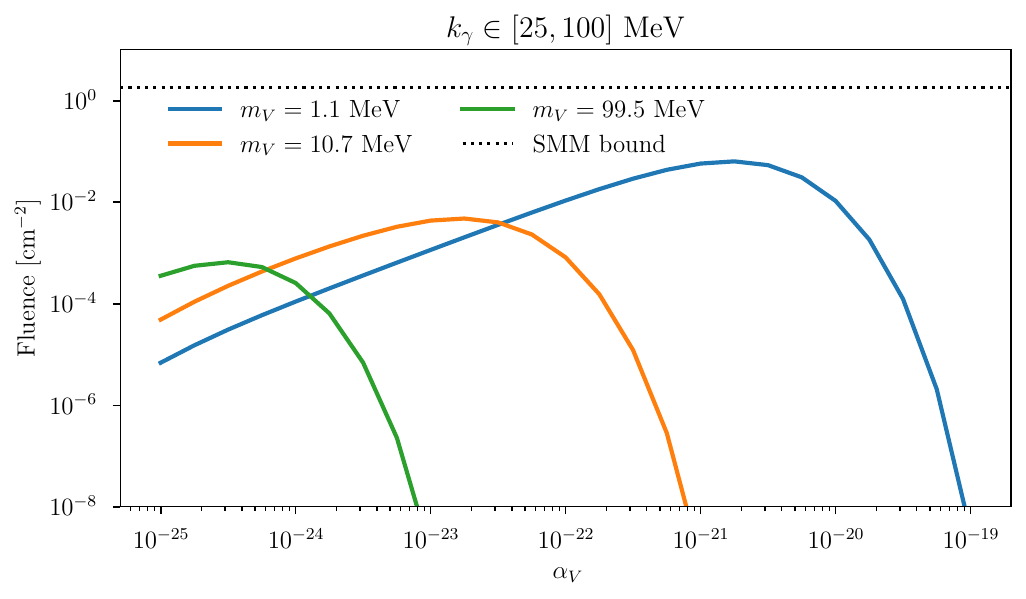}
    \caption{Expected gamma ray fluence from SN1987A (via $V\to e^+e^- \gamma$) as a function of the boson coupling $\alpha_V$ (coloured lines corresponding to different values of $m_V$). We compare these predictions to the limit from SMM (dashed black line). Other gamma ray energy windows yield similar conclusions.}
    \label{fig:smm_fluence}
\end{figure}

Finally, we validate the previous estimates and compute the full result, \cref{eq:fluence_full}, numerically and compared it to the SMM limits. In~\cref{fig:smm_fluence} we show the expected $\gamma$ fluence in one of the SMM channels as a function $\alpha_V$ for three different mass points. We see that the fluence falls well below the experimental limits, and it is numerically consistent with the estimates above. Other channels (photon energy windows) show similar behavior.

We have used a similar calculation (with $\br(\mu\mu)\sim 0.5$) to estimate fluence from $V\to \mu^+\mu^- \gamma$ for $m_V > 2m_\mu$. While some of the above approximations are worse for this channel, we expect them to be sufficient for an order-of-magnitude estimate. We find again the fluence is below the SMM sensitivity in this regime, despite the larger branching fraction. The reason is clear from ~\cref{fig:smm_fluence}: the $\gamma$ fluence from heavier particles peaks at smaller couplings, which also correspond to lower production rates. 

\subsection{Diffuse $\gamma$ Rays from Previous Supernovae}
Gamma rays emitted from $V$ decays of all previous supernovae are constrained by flux limits on extragalactic background light (EBL). We follow \cite{Caputo:2021rux} to estimate this constraint. In the relevant parameter space, $V$'s are cosmologically short-lived, $\tau_V \ll H_0$, enabling us to set the cosmological decay fraction ($f_{\rm D}$ in \cite{Caputo:2021rux}) to 1. Generalizing their ALP result to our approximate photon spectrum, \cref{eq:approx_photon_spectrum}, yields the flux spectrum (in units of $\mathrm{cm}^{-2}  s^{-1} \MeV^{-1}$)
\begin{equation}
  \frac{dn_\gamma}{dk}
  = \frac{\alpha\,\br(e)}{\pi\,k}
    \int_0^\infty dz\;n'_{cc}(z)
    \int_{(1+z)k}^\infty d\omega_z \frac{dN_V}{d\omega_z}(\omega_z),
  \label{eq:diffuse_V}
\end{equation}
where the $V$ number spectrum $dN_V/d\omega$ is defined in \cref{eq:V_number_spectrum}. Here $k$ is the observed photon energy, $\omega_z$ is the $V$ energy at emission, and $n'_{cc}(z)$ is the core-collapse rate per redshift interval per comoving volume. The latter quantity peaks at $z\approx 1$, so following \cite{Caputo:2021rux} we use a simple approximation, $n'_{cc}(z) \approx n_{cc} \delta(z -1)$ with $n_{cc}$ a constant, which allows us to carry out the redshift integral:
\begin{equation}
  \frac{dn_\gamma}{dk}
  \approx \frac{\alpha\,\br(e)}{\pi\,k}n_{cc}
    \int_{2k}^\infty d\omega_z \frac{dN_V}{d\omega_z}(\omega_z).
  \label{eq:diffuse_V_simplifed}
\end{equation}
The remaining integral is simply the number of $V$ with energy above $2k$ produced by typical SN. 

Observed EBL fluxes are presented in terms of a slightly different variable 
\beq
k^2 \frac{d\Phi_\gamma}{dk} \equiv \frac{k^2}{4\pi}\frac{dn_\gamma}{dk} \sim 2\times 10^{-3}\;\MeV\mathrm{cm}^{-2}s^{-1}\mathrm{sr}^{-1}
\label{eq:observed_ebl_flux}
\eeq
which is approximately constant for $k$ between $2$ and $200$ MeV. Using the total number of $V$ (for $m_V = 10\;\MeV$ as in the previous sub-section) as an upper bound on the integral in \cref{eq:diffuse_V_simplifed} and setting $n_{cc} \sim 10^{7}$, we obtain and approximate predicted EBL flux of:
\beq
\frac{k^2}{4\pi} \frac{dn_\gamma }{dk} \lesssim 10^{-8}\;\MeV\mathrm{cm}^{-2}s^{-1}\mathrm{sr}^{-1} \left(\frac{\alpha_V}{10^{-20}}\right)\frac{k}{10\;\MeV}.
\eeq
This is safely below the observed flux~\cref{eq:observed_ebl_flux} for couplings below those already excluded by, e.g., high energy neutrino scattering bounds (c.f.~\cref{fig:combined_sn_bounds}). We have verified this approximate calculation by full numerical integration of \cref{eq:diffuse_V_simplifed} for several parameter points.

\section{Fireball formation}
\label{sec:fireballs}
Another type of observational constraint on dark sectors arises from the formation of a fireball during a SN~\cite{Diamond:2023scc,Candon:2025ypl} or NS-NS merger~\cite{Diamond:2021ekg,Diamond:2023cto}. The fireball is a hot plasma outside the SN or metastable NS, which can process $\sim {\rm MeV}$-scale electrically-charged particles into $\sim {\rm keV}$ scale X-rays. The formation of a fireball requires efficient  thermalization of the emitted electromagnetically-interacting particles.

Let us imagine a simplified cartoon in which $V$ emission from the supernova may be thought of as occurring in concentric spherical shells. These shells of $V$ move radially, and convert into shells of daughter particles as the $V$'s decay; we will be concerned with charged leptons $\ell^+\ell^-$. The momenta of the decay products are distributed as a top-hat (flat in longitudinal momentum) and so the leptons travel at different speeds. It then follows that there are shell-crossings of the charged particles, where they may collide and interact. During shell crossings, $\ell^+\ell^-$ scattering can lead to a thermalized electromagnetic plasma provided the scattering rate is efficient. Alternatively, the $p_T$ imparted from the $V$-decay can lead to oblique crossings of adjacent cells; in any case this allows $\ell^+\ell^-\rightarrow \gamma \gamma$ and other processes to proceed.

In the absence of re-scattering, the number of charged particles per unit area of the shell is given by $\rho_\perp(R)= {\rm BR}_{\pm} N_V / (4\pi R^2)$, where $N_V$ is the total number of $V$ emitted and ${\rm BR}_{\pm}$ is the branching fraction into $\ell^+\ell^-$. Multiplying by a typical electromagnetic cross section $\sigma_{\rm EM} \sim \pi \alpha^2/s$, a necessary condition for thermalization becomes $\rho_\perp(R) \sigma_{\rm EM} \gg 1$ (this is derived rigorously in ~\cite{Diamond:2023scc,Candon:2025ypl}). Simultaneously, for the fireball to be detectable with X-ray telescopes, it must form outside of the opaque regions of the SN or the merger remnant. This means that the decay length of $V$
$$\gamma c\tau_V = 3.5\times 10^{9}\;\mathrm{cm} \left(\frac{10^{-20}}{\alpha_V}\right)\left(\frac{10\;\mathrm{MeV}}{m_V}\right)^2\left(\frac{E_V}{60\;\mathrm{MeV}}\right)$$ 
must exceed $R \sim 10^{10}$ cm for type Ic SN~\cite{Candon:2025ypl} or $R \sim 10^8$ cm~\cite{Diamond:2023cto} for the merger.

For SN, a rough order-of-magnitude estimate (using numerically-computed $N_V$) suggests that the thermalization requirement is incompatible with the minimum detectable radius regardless of $m_V$.

In the case of NS-NS mergers, the situation is more ambiguous. We find that for $m_V \geq 2 m_\mu$ the production rate of $V$ is suppressed; additionally, $V$ with these masses have quasi-relativistic velocities which means that they may not overtake the merger ejecta (which can reach speeds up to $0.8c$). The combination of these effects leaves little room fireball formation, but does not entirely preclude it. If new constraints exist in the high mass region, they will be susceptible to the ejecta velocity uncertainties. Evaluation of these potential bounds requires merger-specific density profiles~\cite{Diamond:2023cto}, and we therefore leave it for future work. 

For $m_V\leq 2 m_\mu$, we find that a fireball can plausibly form for $\alpha_V \geq 10^{-20}$, corresponding to $ g_{\mu\tau} \geq 3.5 \times 10^{-10}$. This region is covered by the constraints derived from high-$E_\nu$ scattering (see \cref{fig:combined_sn_bounds}). This can be understood as stemming from a combination of two effects relative to e.g., an axion coupled to photons, or a conventional dark photon: {\it i)} there is a $\sim 10^{-5}$ suppression in the branching ratio to $e^+e^-$ weakening the fireball, and {\it ii)} there is an $O(1)$ branching ratio into neutrinos strengthening the high-$E_\nu$ constraints.  Based on these estimates, we do not expect the fireball to supply additional constraints, but a definitive answer to this question requires detailed analysis.

\bibliographystyle{apsrev4-1}
\bibliography{refs}

\end{document}